# Vocational Training Programs and Youth Labor Market Outcomes: Evidence from Nepal

SHUBHA CHAKRAVARTY[a], MATTIAS LUNDBERG[a],
PLAMEN NIKOLOV☆[bcde], JULIANE ZENKER[f]

☆Lead and corresponding author: Plamen Nikolov, Department of Economics, State University of New York (Binghamton), Department of Economics, 4400 Vestal Parkway East, Binghamton, NY 13902, USA. Email: pnikolov@post.harvard.edu

[a] The World Bank
[b] State University of New York (Binghamton)
[c] IZA Institute of Labor Economics
[d] Harvard Institute for Quantitative Social Science
[e] Global Labor Organization
[f] Göttingen University

Acknowledgements: Data collection for this paper has been supported through the Swiss Development Corporation (SDC), the UK's Department for International Development (DFID), The World Bank's Multi-donor Trust Fund for Adolescent Girls. We thank Siroco Messerli and Bal Ram Paudel for facilitating all logistical aspects of this study. Headed by Madhup Dhungana, New Era Limited provided exceptional support with survey design, survey implementation and data management. At the World Bank, Jasmine Rajbhandary, Venkatesh Sundararaman, and Bhuvan Bhatnagar led the AGEI project. Amita Kulkarni, Uttam Sharma and Jayakrishna Upadhaya coordinated survey activities. Matthew Bonci, Ali Ahmed, Dayne Feehan, Jake Tuckman, Marine Gassier, Pinar Gunes, and Jennifer Heintz provided outstanding research assistance. Constructive feedback was received from Markus Goldstein, Grant Miller, Eric Edmonds, Susan Wolcott, Chris Hanes, Tristan Zajonc, Jose Luis Montiel Olea, Kristian Rydqvist, Barry Jones, Leonard Goff, Solomon Polachek, Sarah Haddock, Niklas Buehren, Joao Montalvao, Jessica Leino, Brooks Evans, and seminar participants at the World Bank (in its Washington, DC and Nepal offices) and the Economics Department at The State University of New York (at Binghamton). Plamen Nikolov gratefully acknowledges research support by The Harvard Institute for Quantitative Social Science, the Economics Department at the State University of New York (Binghamton), the Research Foundation for SUNY at Binghamton and the EU Commission Marie Curie Micro One World Research Grant. Juliane Zenker acknowledges funding from the Growth and Economic Opportunities for Women (GrOW) initiative. All remaining errors are our own.

# I. Introduction

In much of the developing world, unemployment among the youth is extremely high: youths (ages 16-24) comprise 40 percent of the world's unemployed while accounting for only 17 percent of the world's population (United Nations, 2012; ILOSTAT, 2017). High unemployment is not only related to high poverty but also has a strong influence on other important social outcomes: it impacts crime rates (Blattman and Annan, 2016; Fella and Gallipoli, 2014), depression prevalence (Frese and Mohr, 1987), substance abuse rates (Linn et al., 1985), and rates of social exclusion (Goldsmith et al., 1997). Moreover, the low labor force participation of women is particularly pronounced in many regions of the world, resulting in female-specific consequences of unemployment and underemployment, such as low decision-making power in the household and domestic abuse (Majlesi, 2016; Lenze and Klasen, 2016). Therefore, targeting youth unemployment with effective interventions, specifically taking female needs into account, is one of the highest priorities for low-income countries (World Bank, 2013). Although there are numerous determinants for high levels of unemployment and poverty, lack of skills is arguably one of the most important (Heckman et al., 2004; Malamud and Pop-Eleches, 2010). One common policy response in an effort to enhance skill formation among the youth is vocational training programs. To date, more than 700 youth employment programs from 100 countries have been implemented, and more than 80 percent of them offer skills training.[1]

In this paper, we examine one of the largest youth training interventions in Nepal, which serves almost 15,000 poor and disadvantaged young men and women annually by subsidizing skills training and employment placement services. Despite the rapid expansion of skill-enhancement programs across the world, this study is among the few to rigorously evaluate such a program in a low-income country. Youth employment rates (other than in subsistence agriculture and informal activities) are exceptionally low in Nepal. In particular, women face difficulties entering the non-farm sector or finding paid employment. In combination with a comprehensive advertisement strategy the training program was created to specifically target young Nepali women.

We examine the program's labor market outcomes based on quasi-experimental techniques. We use a large panel data set of three consecutive cohorts of applicants to the program. Program eligibility was based on individual scores determined by a standardized application procedure and a course-based threshold score. Therefore, we exploit the individual-assignment scores to implement a regression discontinuity design. Because we find some evidence that the actual individual assigned score was manipulated in practice, we instead use application-form data to reconstruct the underlying score components and generate our own individual-based score. We then use the reconstructed forcing variable as an instrument for training to estimate the program impacts on compliers (LATE estimates) and as a treatment variable to estimate the programs intent-to-treat (ITT) effects.

[1] See Youth Employment Inventory (http://www.youth-employment-inventory.org/)

We report three major findings. First, approximately twelve months after the start of the training program, the intervention generated an increase in non-farm employment of at least 10 percentage points (based on ITT estimates) and up to 31 percentage points for compliers (based on LATE estimates), both of which are heavily driven by women starting self-employment activities inside (but not outside) their homes. The program also generated an average monthly earnings gain of at least 659 NRs (≈9 USD) (based on ITT estimates) and up to 2,113 NRs (≈28 USD) (based on LATE estimates) for women. Second, in contrast to women, men do not significantly gain from the program along the extensive margin of non-farm employment in the short run but do show an increase in earnings conditional on any employment of at least 698 NRs, which suggests that they use the program to upgrade their skills. Third, using a small sub-sample of the initial study population, we find suggestive evidence that 24 months after the start of the program, men have gained along the extensive margin with an increased non-farm (self)-employment rate of at least 19 percentage points, while female medium-term employment and earnings effects stay in a similar range compared to the short-term effects but, possibly due to the lower statistical power, turn insignificant.[2]

Our program impacts stand in stark contrast to the vocational training literature from middle-income countries as these studies generally find small or no effects (McKenzie, 2017). We highlight three explanations of this discrepancy and how they relate to the low-income country context of our study. First, differences in baseline educational attainment suggest that the magnitude of program effects might be very different in low-income countries than those found in middle-income countries.[3] High poverty rates often go hand-in-hand with low educational levels and, to the best of our knowledge, the baseline educational levels in our study sample are the lowest among the existing experimental and quasi-experimental literature.[4] Low baseline human capital implies a relatively steeper increase in output (and earnings). This relationship between low human capital and large marginal increases in earnings may be further pronounced if gender inequality in educational attainment is high (see, e.g., Knowles et al., 2002; Klasen, 2002).

[2] The increase in effect sizes from the short- to medium-term is in line with results of other studies: Lechner et al. (2011) find average treatment effects on the treated in the form of increased employment rates of 10–20 percentage points in West Germany. Also see Kluve (2010), Card et al. (2010), and Ibarraran (2015) who find evidence of larger medium-term effects.

Second, the measurement of outcome categories can also account for differences between results in middle-income and low-income countries. Low-income countries without institutionalized unemployment insurance are often characterized by a large agricultural and informal sector. The majority of individuals are involved in either subsistence agriculture or informal small-scale businesses as formal employment activities are unavailable (Nayar, 2011). Therefore, unemployment, as it is known in high-income countries, is relatively rare, whereas underemployment is very widespread. Comparing effect sizes between low-income and middle- or high-income countries is further complicated by the fact that the program impacts found within the existing literature in middle-income countries is gauged by any employment or formal versus non-formal employment.[5] However, both of these criteria are generally not useful in analyzing low-income country samples.[6] Vocational training in a low-income country is unlikely to yield large employment effects on the extensive margin or on formal employment rates, but can generate large impacts on the intensive margin (hours worked) and on changes in sectorial composition of employment (farm versus informal non-farm employment), which is indeed the case in our study. Our estimated ITT impacts on the probability of any employment (4 percent, insignificant) or wage-employment (3 percent, insignificant) are in fact in line or smaller than the coefficients on wage-employment found in middle-income countries: for example, Attanasio et al. (2011), Card et al. (2011), and Reis (2015) find coefficients that range between 0 and 8 percent.[7] In contrast, our large effects on the probability of self-employment seem unique to the low-income country context. The few studies from middle-income countries that measure entrepreneurial activities usually find no impact on the probability of self-employment or on self-employment earnings (e.g., Attanasio et al., 2011; Maitra and Mani, 2017).

[3] Betcherman et al. (2007) review findings from vocational programs around the world. The authors posit that programs from middle-income countries are generally more effective than programs from high-income countries, thereby suggesting that program effects may be even larger for vocational training in low-income countries.

[4] Based on the program eligibility criteria, very few of the participants had completed higher than a 10th grade education (approximately 12 percent) and a total of 38 percent of individuals in our sample had only primary education or no education. This is dissimilar to other studies from middle- and low-income countries. For instance, Card et al. (2011) report an average of about nine years of schooling at baseline. Alzua et al. (2016) report that only 11 percent of the sample had primary education or lower, while 57 percent of the sample had completed secondary schooling or more. Hirshleifer et al. (2015) report, on average, eleven years of schooling at baseline. Maitra and Mani (2017) report that 45 percent of their sample had completed secondary education at baseline. Adoho et al. (2014) report 46 percent of their sample had attained an education level of 10th grade or above.

[5] Formal employment is often defined as paid employment that is based on a formal contract and possibly comes with social security benefits like health care, pension, and/or injury compensation, etc. (see e.g. Attanasio et al., 2011).

[6] In our study, individuals employed in the formal sector make up only 1.5 percent of our sample at baseline and 2.5 at follow-up. In Adoho et al. (2014), conducted in Liberia, only five to six percent of the sample is wage-employed at baseline, while formal employment is virtually non-existent. In contrast to these findings, Attanasio et al. (2011) report that in their sample, 7 and 12 percent of women and men at baseline are active in the formal sector, respectively; at the follow-up stage, these numbers are 23 and 38 percent for women and men, respectively. Hirshleifer et al. (2015) detect a share of 30 to 43 percent formally employed participants in the control group.

Third, the fact that the impacts we detect operate mainly through female self-employment inside but not outside the house points to an important dimension in how effect channels may differ across countries: social and cultural norms surrounding work. In traditional societies, women may be more confined by gender roles that constrain their involvement in the labor market. Restrictive norms regarding female labor force participation exist in many regions of the world, but they seem to be particularly strong for women in South Asia (World Bank, 2011; Asian Development Bank, 2015; Booth, 2016).[8] Our results suggest that the program enables otherwise underemployed women to earn an income while staying at home – close to household errands and in line with the socio-cultural norms that prevent them from taking up employment outside the house. Further, certain ethnicities or castes may face obstacles in utilizing their labor market potential as a result of discrimination by employers or lower access to education (Banerjee and Knight, 1985; Mainali et al., 2017). If a vocational training program manages to unlock this latent potential by taking the unique needs of particularly disadvantaged populations into account—as has been done in the program studied here—it might unleash large employment and earnings potential, whereas, in the opposite case, a similar program might prove less effective (e.g., Cho et al., 2016).

Our paper makes three important contributions to the existing literature on vocational programs in developing countries. First, our paper provides evidence that large program impacts of vocational training programs – particularly in the low-income context – are possible, despite most of the existing evidence from middle-income countries suggesting otherwise.[9,10] Based on the experience of middle-income countries, only Maitra and Mani (2017), Reis (2015) and Alzua (2016) find positive impacts on the probability of any employment and any earnings in India, Brazil, and Argentina, respectively. On the other hand, Honorati (2015), Card et al. (2010), Attanasio et al. (2011, 2015), Ibarraran (2015), Hirshleifer et al. (2015), Acevedo et al. (2017), Diaz and Rosas (2016), and Galasso et al. (2004) find either mixed, muted, or no impacts at all from vocational programs on various labor market outcomes.[11] That large effects may be particularly pronounced in low-income contexts is confirmed by two studies conducted in Liberia and Uganda[12]: Adoho et al. (2014) randomly assign a similar intervention like the one studied here and detect an increase of 47 percent in non-farm employment and 80 percent in earnings among young Liberian women. Similarly, Bandiera et al. (2017) find positive impacts on income generating activities of 48 percent (which were almost entirely driven by self-employment), but no positive impacts on wage-employment in Uganda. Our second contribution to the existing literature relates to the pattern of different returns to vocational training between women and men. Although Blattman and Ralston (2015) point to a stylized fact that proposes vocational training has higher returns for women, McKenzie (2017) reviews recent vocational training programs in low-income and middle-income countries and, in fact, argues that previous studies, which formally test for equality by gender, can either not reject similar impacts for men and women, or have found significantly higher impacts for men.[13] In stark contrast, our study does formally test for equality by gender and it unambiguously shows robust evidence that vocational training in our context yields higher returns for women. We highlight that our results are likely driven by the socio-cultural norms in Nepal, which shape gender roles that identify women with more restrictive characteristics and capabilities in the labor market compared to other country contexts in the training literature (e.g., Latin America). This exemplifies that generalizing heterogeneous impacts of policies such as the one investigated here should (if at all) be done very cautiously, carefully considering the relevant effect channels. Finally, our study underscores that measuring a wider range of employment outcomes, such as self-employment, may be necessary to comprehensively study the impact of active labor market programs. Even though formal employment per se is not affected by the vocational training in our context, we are able to identify some short run effects on women's self-employment that less comprehensive labor market surveys of previous studies may have missed.

[7] Overall, these studies provide skepticism regarding the cost-effectiveness of training programs (Almeida et al., 2012). In contrast, the Liberia experiment, mentioned above, cost between $1,200 and $1,650 per beneficiary; although this is relatively high, the returns to self-employment training are also sufficiently large to recoup this cost within three years (Adoho et al., 2014).

[8] Only one other quasi-experimental study in South Asia examines the impact of training programs on employment and earnings. Maitra and Mani (2017) evaluate a training program in stitching and tailoring offered to young women in New Delhi. They find that program participation increased employment by more than 5 percentage points, self-employment by almost 4 percentage points, and any employment by 6 percentage points. The program increased number of hours worked by approximately 2.5 hours. However, the smaller effect sizes are based on ITT estimates, which are based on much lower compliance rates (56 percent) compared to our study (70 percent). Also, initial levels of wage-employment and education are higher in India as Nepalese socio-cultural norms regarding gender roles may be less restrictive in New Delhi compared to rural Nepal where most of our courses take place.

[9] We use the World Bank income classification of countries from http://databank.worldbank.org/data/download/site-content/CLASS.xls

[10] Most of the vocational training literature on similar programs in high- or middle-income countries finds low or insignificant effects (Card et al., 2010; Kluve, 2010; Dar and Tzannatos, 1999).

[11] Other authors document similarly sized impacts based on related active labor market policies such as wage subsidies (Galasso et al., 2004; Groh et al. 2016; Levinsohn et al., 2014)) or search and matching assistance (Adebe et al., 2016a; Franklin, 2005; Adebe et al., 2016b; Jensen, 2012; Groh et al., 2015; Dammert et al., 2015; Beam, 2016; Abel et al., 2016; Bassi and Nansamba, 2017)

[12] Rigorous evidence of one other vocational program based in low-income countries exists: In Malawi, Cho et al. (2016) find no impacts on hours worked and no impacts on other employment outcomes. Yet, the authors explain that an unsatisfactory tailoring of the program to the needs of the target group in combination with large numbers of drop-outs are likely responsible for the low effects.

[13] To the best of our knowledge, considering the rigorous evidence on vocational training programs, only Alzua et al. (2016) formally test for gender-disaggregated effects *and* find a difference: the authors document larger impacts for men.

The remainder of the paper is structured as follows. Section II provides background information on Nepal's labor market and details Nepal's Employment Fund training program and the intervention design. Section III describes the data, sampling strategy and sample characteristics. Section IV presents the study design, and Section V describes the empirical strategy. Section VI presents the results, and Section VII provides various robustness checks. Section VIII concludes.

## II. Background

### A. The Labor Market in Nepal

Nepal's economy ranks among the world's poorest. In 2010, a quarter of the country's population lived below the national poverty threshold. The Nepalese economy is characterized by lack of formal sector jobs, a large informal sector, and wide-spread underemployment (affecting approximately half of the workers in the younger age groups), all of which contribute to very high poverty rates (ILO, 2004; Central Bureau of Statistics, 2009).[14,15] Based on the Nepal Labour Force Survey 2008, 74 percent of laborers work in the agricultural sector while 64 percent work in subsistence agriculture. Women are particularly underrepresented in non-agricultural employment. Although the Nepal Labour Force Survey 2008 reports a labor force participation of women (80.1 percent) that is similar to that of men (87.5 percent), approximately three quarters of employed women work in subsistence agriculture (compared to 52.9 percent of men), whereas only 26.2 percent are engaged in paid work (compared to 73.9 percent of men).

[14] Although no labor survey specifically collects informal sector data, some economic measures point to the size of the informal sector constituting more than two-thirds of the economically active population in Nepal (Suwal and Pant, 2009). One of the reasons for Nepal's bad economy and its underdeveloped formal sector is the political turmoil the country experienced in the last two decades: a peace agreement between the government and the Maoist insurgency, an interim constitution promulgated in 2007, the 2008 declaration of a democratic republic, new and rising ethnic political movements, and a democratically elected Constituent Assembly in 2013.

[15] Nepal's 2008 labor force survey (Central Bureau of Statistics, 2009) shows that 21 percent of the individuals working less than 40 hours per week are in fact able to work more and are therefore underemployed. Shortage of employment opportunities has generated a migration flow to urban areas and migration to other countries, especially to the Gulf States (Lokshin and Glinskaya, 2009). Nepal's net migration has registered an outflow of migrants near half a million for the period 2008 to 2012. During the same period, Nepal's net migration rate has exceeded the same number for the overall migration rate of South Asia.

Moreover, monthly earnings of female paid employees are much lower on average (NRs 3,402 versus 5,721 for men) (Central Bureau of Statistics, 2009).

*Low skills and high youth unemployment.* The country's labor market outcomes can, to some degree, be attributed to low levels of human capital accumulation. According to the Nepal Labour Force Survey 2008, the total literacy rate for individuals of ages 15 years and above is 55.6 percent (70.7 percent for men and 43.3 for women). Only 22 percent of women and 29 percent of men have more than a primary education, and approximately half of the Nepalese have no formal schooling at all (Central Bureau of Statistics, 2009). These numbers are staggeringly low even in comparison to the human capital indicators of other South Asian countries and the rest of the developing world. For instance, in Latin America, where most of the available experimental evidence on the effectiveness of training programs in developing countries comes from, approximately 75 (50) percent of the population of age 15 (20) or above had completed primary (lower secondary) education in 2004 (UNESCO, 2007). Moreover, lack of vocational skills is predominant among young Nepalese. Around half a million young people join the labor force each year, the vast majority of them being unskilled. Although the market demand for skills is high, access to vocational training is limited, particularly among the poor and disadvantaged groups.

*Cultural and Social Norms Regarding Female Employment.* A combination of low educational attainment and restrictive norms towards marriage, childbirth, and household duties generate multiple constraints for young women who wish to enter the labor market. These constraints are reflected in the different reasons that young men and women report for not participating in the labor force: although the majority of economically inactive men between ages 15 and 29 report that they are attending school (85.3 percent), only 43.9 percent of women of the same age group report school enrollment as their major reason. Instead, 41.6 percent of economically inactive women between ages 15 and 29 and even 80.1 percent of women between ages 30 and 44 report to be economically inactive because they have to engage in (unpaid) household duties, while virtually none of the men in these age groups state household duties as a reason for being absent in the labor market (Central Bureau of Statistics, 2009).

In addition, being bound to their households and families because of gender roles, women also face three broad gender-based employment barriers, which further complicate their economic well-being: restricted mobility, cultural norms, and societal norms regarding gender expectations for certain occupations. Largely influenced by Hindu philosophy related to men and women's positions in society, Nepal's socio-cultural practices differ by caste and ethnicities. For example, related to the first employment barrier, in southern Nepal (Terai and Madhesh) women are frequently confined to the household and are unable to travel outside of their immediate community for any work. Various proscriptions based on the Madhesi culture restrict females from leaving their homes. The second gender-related barrier relates to cultural norms that prohibit women from interacting with men other than their family members. For example, in conservative parts of the Terai districts, females who are trained as barbers are not employed due to cultural prohibitions against women touching men. The third gender-based labor market barrier relates to women who choose to train and work in traditionally male-dominated occupations. In various typically male-dominated occupations – for example, computer and television repair, auto body making and construction work – customers frequently express doubts regarding the quality of skills of women assigned to such repair jobs. As a result of the societal difficulties women face, they remain engaged in unpaid, home-based labor to a large extent. Approximately 80 percent of the unpaid family labor force is female (Central Bureau of Statistics, 2009).

In summary, the low educational levels and the shortage of skill training opportunities suggest that the return to obtaining additional training – especially for women – may be particularly high in Nepal. With that said, training for women may only be successful if the program takes female-specific needs into account so that it can enable them to expand their economic activities while adhering to their social roles and the cultural norms of their communities.

### B. The Nepal Employment Fund

Started in 2008, the Employment Fund (EF) is one of the largest skills training initiatives in Nepal. The program provides vocational training and placement services under a unique governance structure in cooperation with local training providers. The EF program subsidizes short-term market oriented skill training in combination with other services for disadvantaged young women and men. The fund's objective is to place trainees into gainful employment upon training completion. Each year, the EF sponsors about 600 to 700 training courses. The EF-subsidized courses are announced publicly in local communities with the intent to encourage potential candidates to apply. The applicants are then selected by a standardized procedure based on eligibility criteria. Available seats are restricted due to the limited capacity of training providers. As of 2010, the program operates at scale nationwide and covers 54 districts, providing training for over 65 occupations and has expanded in the consecutive years. Table 1 provides an overview of the total number of training events, trainees, and training providers for the time of our study period (2010-2012).

*Program components.* Admission to the program offers the trainees a bundle of services, where the core components are technical training (including certification) and job-search-assistance. Training courses in technical skills vary across a wide range of trades (e.g., incense stick rolling, carpentry, tailoring, welding, and masonry) and last from four weeks to three months. Each trainee is encouraged to complete a skills certification test offered by the National Skills Testing Board (NSTB). Upon completion of the classroom-based training, the EF emphasizes job placement services. Once the first training phase is completed, training providers are required to link trainees to trade-specific employers for six months of paid on-the-job-training. Providers often use their trade-specific networks of trainers and employers to find suitable internships for their graduates. Through the internship, trainees obtain immediate work experience as they apply their learned skills in the market and, subsequently, strengthen their social capital and contacts with potential employers. In addition to the core components of the program, all females receive 40 hours of life-skills training (started in 2010 and fully implemented in 2011). The forty-hour training curriculum covered topics such as negotiation skills, workers' rights, sexual and reproductive health and discrimination response. A subset of trainees also received a short course in basic business skills.

*Outcome-based payments.* Training providers are rewarded for their services in three installments based on a set of pre-determined outcomes. A provider qualifies to receive full payment – i.e., the full price of the training and services provided plus a bonus based on whether the trainee belongs to a vulnerable group – by a set contract between the EF and the training provider. The contract stipulates that all accepted trainees must successfully complete their skills training and sit for a skills test given by the Nepal National Skills Testing Board (NSTB). When trainees complete the test, providers obtain their first installment (40 percent of total payment). Upon the exam completion, providers are expected to ensure that graduates remain continuously employed for the next six months and that they earn above 3000 NRs per month ("gainful employment"). The EF verifies three months and six months after training completion whether or not trainees are gainfully employed. If the verification process is successful, the training provider obtains the second and the third installments (25 percent and 35 percent of total payment, respectively).[16] The cost for training and employment services is pre-financed by the providers and is reimbursed to them only when they accomplish the outcomes in their contract. Therefore, training providers bear the risk of losing their investment if they are unsuccessful in training and placing their trainees accordingly.

*Targeting disadvantaged groups.* Three factors comprise the eligibility criteria for all EF-sponsored training programs: age (from 16 to 35), education (below SLC[17], or less than ten years of formal education) and low self-reported economic status. Only applicants who meet all three criteria were viable for being short-listed in the admission procedure. Furthermore, and as mentioned above, providers receive a bonus payment for successfully training and placing candidates who belong to particular disadvantaged groups.[18] The bonus payment is calculated as a percentage of the full cost of training and services provided and is issued proportionally together with the three installments described above, as long as the particular requirement for each installment is fulfilled. A provider receives a bonus payment of 40 percent of the base cost of training for a man who is poor and 50 percent for a man who is poor *and* belongs to a disadvantaged group. A provider further receives a bonus payment of 70 percent of the base cost of training for a woman who is poor and 80 percent for a woman who is poor *and* belongs to a disadvantaged group.[19]

In 2010, the EF, partnering with the World Bank, made additional efforts to specifically target young women aged 16 to 24. Training under the Adolescent Girls Employment Initiative (AGEI) proceeded in the same way as it did for other EF trainees, except that certain events had been flagged in advance as likely to attract female trainees. In addition to regular training course advertisement, the EF sponsored radio and newspaper ads specifically geared towards young women. Many of these ads specifically encouraged women to sign up for traditionally male trades, such as mobile phone repair, electronics, or construction.

---

[16] Employment status of a randomly selected sample of graduates is verified by EF field monitors.

[17] The School Leaving Certificate (SLC) is obtained after successfully passing examinations after the 10th grade. To be eligible, EF applicants must have not taken, or not passed, their SLC exams. This criterion has been loosened for some trades starting in 2012.

[18] Disadvantaged groups are defined by the EF as people belonging to the Dalit community, ex-combatants, internally displaced, widows (only women), disabled, HIV/AIDS infected, and formerly bonded laborers.

[19] Poverty is defined as less than six months of food sufficiency for farm households or less than 3,000 per capita family income, from non-farm based income.

## III. Data

### A. Sampling

We used two primary sources of data. First, we used data from training application forms and the selection procedure of EF-sponsored training that covered three consecutive cohorts of applicants (from 2010 to 2012). Second, we conducted individual (applicant) and household surveys with two rounds of data collection for each cohort. For the 2010 cohort, a second follow-up was conducted on half of the cohort. Figure 1 shows the survey timeline.

[Figure 1 about here]

Sampling into this study included a combination of stratified, random, and convenience sampling and was done in two consecutive steps. For each cohort, the first step consisted of a selection of training events, and the second step consisted of selecting individuals according to standardized procedures. The event sampling-frame for this study consisted of all training events from the universe of the EF funded trainings that occurred between January and April of each year. Events were grouped into clusters of close-by districts before sampling for survey administrative reasons. We then randomly sampled up to 15 district clusters in each of the three years, respectively. Furthermore, from the list of training events that took place in these district clusters, we randomly selected 20 percent. Because of the focus on young women in this study, events that were likely to include more young women (identified by training providers) were purposely oversampled in 2011 and 2012. In 2010, because a complete event listing was not available in advance, the events were not chosen randomly but by convenience, based on scheduling and accessibility. Table 1 details the resultant sample of events for the three cohorts.

[Table 1 about here]

The 2010 sample comprised 65 events across 12 districts. The 2011 sample comprised 182 events, of which 113 events were dropped from the baseline survey, either because the survey team could not reach the event on the day of applicant selection (usually due to weather conditions) or because the event was not "oversubscribed".[20] The remaining 69 events in 28 districts were included in the 2011 baseline sample. In 2012, 85 out of 112 sampled events covering 26 districts were included in the study sample. Figure 2 depicts the study areas by survey.

[Figure 2 about here]

To sample applicants, a survey team visited each sampled training event on the day when applicant selection happened. The sampling of applicants was based on the standardized interview procedure that was used to determine training assignments. During the assignment procedure (which we will describe in more detail below), applicants received scores in five different categories that were added up to form a total score. A ranking sheet was then used to list applicants from the top scorer to the bottom scorer and indicated the threshold (i.e., minimum score) for admission to the course. The individuals we study comprise a subset of the ranked individuals -- those who fell in the range of 20 percent below or above the threshold. Immediately following the sampling of applicants but before the results of the selection process were announced, a baseline survey was administered.

[Table 2 about here]

The sampling procedures resulted in a study sample of 4,677 individuals across all three cohorts at baseline, see Table 2. For the first follow-up surveys, we were able to track and successfully interview 88 percent of the baseline survey respondents, yielding a panel for analysis of 4,101 individuals.[21] Because training courses vary in length from one to three months, the follow-up survey examines outcomes nine to eleven months after the end of the training. The EF itself conducts follow-ups with a sample of participants up to six months after the training to verify employment and earnings. This is also the time point when providers receive their last payment installment. Hence, our first follow-up survey occurs three to five months after the training evaluation and the treatment group's last contact with the program.

---

[20] Oversubscription was necessary to obtain a sufficiently large "quasi"-experimental control group as detailed in the description of the applicant sampling below. The survey team was instructed to drop the event from the sample if there were not at least three rejected candidates who could be sampled for the control group (i.e., at least three candidates that fell within 20 percent of the threshold score).

[21] The reasons given for loss to follow-up for the 2010 and 2011 cohorts include: inability to track the household (11 percent), no one in the household during multiple visits (15 percent), refusal (8 percent), respondent migrated for work within Nepal or abroad (8 percent), respondent migrated after marriage (10 percent), or other (40 percent).

### B. Sample Characteristics

We present an overview of baseline characteristics in Table 3. Eligibility to apply for EF-sponsored training courses was restricted by age, education level, and poverty status; therefore, individuals in our sample are on average young, low educated and relatively poor. For the pooled sample (i.e., 2010-2012 cohorts), the study population has an average age of 24 years and is 63 percent female. A total of 38 percent of individuals have either have no education or only primary education. Approximately 61 percent of the sample engaged in some income-generating activity in the month prior to the survey. When we restrict to non-farm income-generating activities, the employment rate falls to 30 percent. The average number of hours worked in the month previous to the survey was 69. At baseline, the average monthly earnings were 1,269 NRs in the month prior to the survey (equivalent to about 17 USD). This figure appears low as it represents the average earnings over the entire study population of 4,677 individuals, including those with zero earnings. Earnings conditional on any income generating activity were 2,082 NRs. Only 18 percent earned more than 3,000 NRs per month, a level deemed to represent "gainful" employment. Furthermore, only 18 percent of the sample was already engaged in the same trade as the training to which they applied, denoted as "trade-specific income-generating activity (IGA)", indicating that a significant minority of applicants had been looking to upgrade existing skills.

[Table 3 about here]

Generally, women have lower *paid* employment levels and earnings at baseline. Forty-seven percent of women engage in activities *inside* the house that yield some income (e.g., self-employment activities), while only 36 percent of the women engage in *paid* activities *outside* the house. In contrast, 59 (69) percent of men engage in *paid* activities *inside* (*outside*) the house.

Also, men (69 percent) are more likely than women (56 percent) to carry out *unpaid* work *outside* the house (e.g., helping relatives); however, more women carry out *unpaid* work *inside* the house (e.g., household chores, child care). Almost all women (94 percent) work in the household without pay for at least five hours a week, whereas this is only true for 61 percent of men. Furthermore, 55 percent of the women work more than 20 hours per week inside the house without pay, which is only true for 12 percent of the men in our sample.

## IV. Study Design

Admission to the program was based on a calculated score for each applicant and coupled with a course-specific threshold score. For each course, applicants with scores above the threshold were assigned to the training program, whereas applicants whose scores fell below the threshold were not assigned to the program. To form a sufficiently large quasi-experimental control group, training providers were advised to shortlist at least 50 percent more candidates than the number of spaces available in the training event. The assignment procedure followed streamlined guidelines, including a detailed scoring rubric, instructions for ranking the shortlisted candidates by score, and selecting the top-scoring candidates for admission to the program.

The individual score used in ranking candidates consisted of five sub-scores based on: (a) applicants' trade-specific education (prerequisite, 15 points)[22], (b) applicants' economic status (up to 20 points), (c) applicants' social caste, gender, and special circumstances (up to 25 points), (d) development status of applicants' district of origin (up to 10 points), and (e) a score determined by a selection committee during an interview procedure (up to 30 points). Sub-scores for the first four components were determined based on the information each applicant provided in his or her application form (shown in Figure A1 in Appendix A). The application form lists the exact questions upon which the distribution of scores was based. Tables A1 and A2 show how these criteria were converted into numeric scores within each of the four categories. Based on the aggregated four-component score, candidates were short-listed and invited for an interview. The fifth sub-score was determined based on an interview with a three- to five-member committee comprised of representatives from the training institution and potential employers. Moreover, representatives of the survey firm and/or the donor institutions (e.g., EF, World Bank, etc.) were usually present to observe the procedure. The selection committee jointly decided on the fifth sub-score by assessing the candidate's commitment, motivation, attitude, aptitude, and clear vision for employment and enterprising. Eventually, the selection procedure yielded a total score for each individual by summing across the five components. Possible total scores ranged from 0 to 100.

[22]Applicants had to fulfill course specific prerequisites (e.g., literacy, certain trade-specific knowledge or experience) to be eligible for the ranking procedure. If an applicant did not fulfill these prerequisites, he was then removed from the selection procedure. If applicants fulfilled the prerequisites, they then received 15 points as their first score-component. In exceptional cases (approx. 9 percent of the sample) this rule was not adhered to, and candidates received 0, 5, or 10 points.

Admission in each course was then based on candidates' decreasing rank; available seats were assigned starting with the top-scorer. Therefore, the threshold in each course was based both on the distribution of candidates' scores and on the number of pre-determined available seats. Because the distribution of scores and the number of seats determine the selection process, the threshold score varies for each course, something we take into account in the empirical strategy described in Section V. Figure A2 displays a sample ranking form used by training providers.[23] Although eligibility for training, based on the actual score, influenced the likelihood of training course enrollment, individual assignment to training was not automatic as it was originally envisioned for two main reasons: non-compliance and manipulation of the assignment procedure. We discuss the implications of these two factors in the remainder of this section.

As we examined the compliance with the selection process, we found that approximately 30 percent of the group assigned to the program did not take-up the training opportunity (possibly due to taking up other training or employment opportunities), whereas 32 percent of the non-assigned applicants did. These take-up numbers can be explained by a simple process in which seats not taken by individuals assigned to training were then given to individuals who were not originally assigned to training but rather next in line based on the ranking form. In the regression discontinuity estimation setup we use, an important step is to examine whether and how the probability of treatment jumped at the threshold. To account for the course-specific threshold, we subtract each individual's assignment score by the course-specific threshold score, whereby we obtain a standardized relative score around the cut-off of zero. We then plot the probability of treatment against the relative assignment score. Figure 3's left graph shows the results of this procedure. The plot reveals a clear jump in the probability of program participation at the cut-off. As expected, the jump is less than one, a fact that we incorporate in the estimation strategy we follow.

[23] Data in each column approximately represents the distribution of the respective component in the full sample. In this example, 15 seats were available and the score of the 15th ranked person on the list was 73. Hence, in this example, 73 would be the threshold score for this particular course.

[Figure 3 about here]

To assess possible manipulation of the admission procedure, we plot the density of the relative score (shown in the left graph of Figure 4). The plot reveals discontinuities in the distribution of the score around zero, which suggests that candidates' scores may have been manipulated to shift certain individuals across the assignment cut-off. In our scenario, such *precise* manipulation of the score was virtually impossible for candidates themselves, as it would have required access to the ranking form after the official selection procedure was completed and the course-specific threshold score was determined.[24] In contrast, providers may have had the opportunity to precisely alter the scores of those candidates who seemed favorable to them.[25] Although the selection committee included several persons from different interest groups and was designed to avoid this type of manipulation on a large scale, we cannot rule out the possibility that providers were able to manipulate the ranking sheet after the official selection procedure was completed. Because possible manipulation of the score can bias our estimates, we specifically address this manipulation issue by reconstructing the assignment score, which we describe in more detail below.

[Figure 4 about here]

[24] Lee and Lemieux (2010) distinguish between *precise* and *imprecise* manipulation. While applicants were certainly able to manipulate the information they gave in the application form, aiming to raise their score, the forms were filled out long before the course threshold was determined, which only happened once all candidates were interviewed, which, as mentioned earlier, occurred after the application forms were submitted to the providers. Applicants' control over their score was therefore *imprecise*, which is actually what sorts them randomly around the cut-off. It is, therefore, not a threat to internal validity and not likely to have caused the discontinuity in our graph (Lee and Lemieux, 2010).

[25] Based on the above described payment scheme, providers had a motive to select candidates for the program who seem most employable or most disadvantaged. Also, favoritism or bribes from otherwise rejected candidates may have played a role in the manipulation.

## V. Empirical Strategy

To estimate program effectiveness on labor market outcomes, we use a non-parametric regression discontinuity strategy by running local linear regressions. Specifically, we estimate local average treatment effects (LATE) for the people who comply with the assignment status, intent-to-treat (ITT) effects, and heterogeneous effects by gender and trade within these two frameworks. In the following section, we describe the estimation strategy, the bandwidth selection, and how we reconstruct the assignment score.[26]

### A. Treatment Effect Estimators

*Local Average Treatment Effects (LATE).* To address imperfect compliance to treatment, we employ a fuzzy regression discontinuity set-up similar to the one proposed by Hahn, Todd, and Van der Klaauw (2001). We run the following first-stage equation:

$$D_i = \beta_0 + \beta_1 T_i + \beta_2 X_i + \beta_3 (X_i - t_c) + \beta_4 T_i (X_i - t_c) + \epsilon_i \, , \qquad (1)$$

where the treatment dummy $D_i$ indicates whether or not an applicant $i$ has received training, and $T_i$ is the excluded instrument, specifying whether an applicant has been assigned to training (i.e., whether the absolute assignment score $X_i$ of the applicant is greater than or equal to the threshold score $t$ of the respective course $c$ he or she applied to). Furthermore, the forcing variable $(X_i - t_c)$ is the applicant's relative assignment score (i.e., it is the difference between an applicant's absolute assignment score and the threshold score $t_c$ of the course). The predicted values of $D_i$ are then used to run a second stage equation:

$$Y_i = \gamma_0 + \gamma_1 D_i + \gamma_2 X_i + \gamma_3 (X_i - t_c) + \gamma_4 D_i (X_i - t_c) + \mu_i \, , \qquad (2)$$

where $\gamma_1$ captures the local average treatment effect or the treatment effect for the compliers, and coefficients $\gamma_3$ and $\gamma_4$ represent the different slopes of the linear regression line left and right of the cut-off, respectively. In both equations, we have added the absolute assignment score as a control variable to account for the heterogeneity in the cut-off values across courses.[27] It is important to note that the LATE estimate is not necessarily equal to the population average treatment effect, as it is based only on those candidates who comply with program assignment. In our sample, it is likely that compliers actually have higher returns to additional education compared to the average individual, an assumption that we further explain in the results section below. We interpret the LATE estimates as the upper bound estimates of the program impacts.

[26] We follow the practical guidelines for regression discontinuity designs in Imbens and Lemieux (2008) and Lee and Lemieux (2010).
[27] Based on suggestions in Cattaneo et al. (2016), we ran a graphical analysis that aggregates courses by the absolute threshold value to fully exploit all the information available by our multi-cutoff setup and explore how treatment effects may vary based on this on this (see Figure A3 in Appendix

*Intent-to-treat effects (ITT).* The overall program effect (regardless of compliance) is of policy relevance. In an attempt to deal with the issue that the complier population might differ from the full sample, we employ an intent-to-treat regression discontinuity set-up in which we treat assignment to training as the treatment variable. We estimate the reduced form equation:

$$Y_i = \delta_0 + \delta_1 T_i + \delta_2 X_i + \delta_3 (X_i - t_c) + \delta_4 T_i (X_i - t_c) \ + \eta_i \ , \qquad (3)$$

where $T_i$ is again an indicator that represents whether an applicant has been assigned to training or not. The coefficient $\delta_1$ can be interpreted as the non-parametric local intention-to-treat effect (Lee and Lemieux, 2010) or the effect of *training assignment* on outcomes. This effect is likely lower than the population average treatment effect in our scenario because several candidates in the assigned group have not been trained, while several people in the non-assigned group have received training—likely biasing the estimate, $\delta_1$, towards zero. We, therefore, interpret the ITT effects as the lower bound estimates of program impacts.

*Heterogeneous local average treatment effects (HLATE).* Because treatment heterogeneity has important implications for eliciting the mechanisms through which the program operates, we estimate heterogeneous local average treatment effects (HLATE) based on the framework proposed by Becker et al. (2013). In particular, we estimate a two-stage procedure similar to the one described above with a second stage represented by the equation:

$$Y_i = \gamma_0 + \gamma_1 D_i + \ \gamma_2 X_i + \gamma_3 (X_i - t_c) + \gamma_4 D_i (X_i - t_c) + \gamma_5 H_i + \gamma_6 H_i D_i + \mu_i \ , \qquad (4)$$

A). Unfortunately, our sample size is not large enough to draw robust conclusions from this analysis as the confidence intervals in the graphs are relatively large. However, some of the graphs do suggest that the program may have been less effective in courses with relatively higher thresholds (i.e., courses that are likely to be located in particularly poor areas or frequented by particularly poor or disadvantaged applicants).

where $H_i$ is an indicator for the subgroup. In the first stage, we use the predicted probability of training and its interaction with the subgroup indicator as instruments for $D_i$ and $H_i D_i$.

Additionally, we estimate heterogeneous ITT effects by adjusting Equation (3) to include the subgroup dummy $H_i$ and its interaction with the assignment indicator.

## B. Determining Bandwidths

For choosing the optimal bandwidth we follow Ludwig and Miller (2007) who suggest a cross-validation procedure to find the optimal balance between precision and bias.[28] The cross-validation procedure chooses relatively large bandwidths in our sample, which in some cases include almost all candidates. We provide plots of the cross-validation functions for all our main outcomes in Figures A4 and A5 of Appendix A. In a robustness section below, we further investigate the sensitivity of our estimates to the choice of bandwidth.

## C. Reconstructing the Assignment Score

In an ideal case, we can examine the effect of training provision on outcomes by using the individual scores assigned by the providers during the interview procedure. The discontinuity in training assignment induced by the threshold score should theoretically generate an exogenous change in the probability of training, holding individual characteristics constant. However, and as mentioned previously in Section IV, we have a reason to believe that training providers were influencing the assigned scores – possibly in response to the payment structure, which rewarded training completion and trainee placement over drop-outs and non-placed trainees. Therefore, manipulating the individual scores is likely to be related to unobserved individual characteristics, and, therefore, likely to bias the estimates of interest (McCrary, 2008). Which direction this bias takes is not obvious. It is likely that applicants who seemed particularly employable were favored by providers, in which case our estimates would be upward-biased. On the other hand, it is also conceivable that providers favored disadvantaged groups, as successfully training those groups was also incentivized with higher final payments. Finally, we cannot rule out the possibility that manipulation may have been the result of bribery or favoritism toward friends or relatives in which case the direction of possible bias could go in either direction. To determine the size and the direction of the potential bias, we run balancing tests on relevant characteristics at the baseline using the LATE and ITT specifications described in Section V. Specifically, we examine the balancing of the outcome variable of interest at baseline in response to potentially influential characteristics such as age, education, gender, and ethnicity, which are likely to determine labor market outcomes. We report these balance tests in Table 3. The difference tests reveal that the initial assignment, based on the original scores, does not perform very well in balancing relevant covariates or labor market potential at baseline. Assigned individuals are more likely to be male, less likely to be of Dalit ethnicity, more likely to have engaged in a non-farm wage-employment activity in the past month, more likely to have worked more monthly hours, and more likely to exhibit higher initial earnings. In accordance with the incentivized payment structure, providers seem to have shifted those candidates across the threshold who appear to have been more employable. Moreover, given the imbalances of gender and Dalit ethnicity, which are in contradiction with the EF incentives to focus training on vulnerable groups, it seems possible that providers may have used score manipulation as a risk reduction or risk diversification strategy. Overall, we conclude that using the original score will most likely bias our results upwards—for the estimation of treatment effects—as a result of its manipulation.

[28] We apply the user-written program *rdbwselect* provided by Calonico, Cattaneo, and Titiunik (2014) to estimate the cross-validation functions that determine our bandwidths.

To overcome this challenge, we follow the approach by Miller et al. (2013) who reconstructed the 'actual' individual-specific score from survey data. Currie and Gruber (1996a, b), Cutler and Gruber (1996), and Hoxby (2001) also follow this approach. As we cannot exactly be sure which of the sub-scores have been subject to manipulation by the training providers, we reconstruct all five of them and later aggregate them to obtain a new total score. We use data from candidates' original application forms to assign three out of five sub-scores. As described above, this data is necessarily free of *precise* manipulation, as the forms were filled out long before the ranking sheet and the course thresholds were determined. Figure A1 in Appendix A shows the section of the application form that contains the relevant applicant information. We assign points based on this information as well as the exact scoring rubric used for the original score (see Table A1 and A2).

For the remaining two score components, we follow two different strategies. The first sub-score refers to the applicants' trade-specific education and was initially meant to be 15 for all short-listed candidates. Usually, if candidates did not fulfill the course-specific education prerequisites, they were not eligible for short-listing and immediately rejected. However, in exceptional cases (approximately nine percent of the sample) this criterion was not adhered to and instead applicants received 0, 5, or 10 points. We, therefore, reconstruct the first component based on an OLS specification that regresses the original first sub-score on candidates' general and course-specific education attainments, each interacted with a set of course dummies. To avoid the estimation bias induced by the aforementioned manipulation, we remove all candidates from the model who fall within five index scores of the cut-off (where most of the manipulation took place).[29] We then predict the outcomes (including all candidates) and round them to values that are factors of five in order to reflect the original distribution of the first sub-score.

For the fifth score component, the selection committee was supposed to assess employability and non-cognitive qualities of the applicants to rate their overall probability to successfully complete the program. If *precise* manipulation was applied to an applicant's fifth sub-score, it was likely carried out vis-a-vis the sum of the other four sub-scores, which were available at the time of the interview. Given the scoring rubric in Table A1, candidates who are better educated, less poor, less disadvantaged, and from a more developed district are ranked relatively lower, but may have higher potential to be successful in the labor market after training completion and, therefore, might be the more interesting candidates for the providers. If this is indeed the case, the incentive for manipulation would be positively correlated with the first score component and negatively correlated to the following three sub-scores. In other words, the higher (lower) the first (second, third, and forth) component-score, the higher the incentive is for the provider to secretly add points onto the applicant's fifth score in order to shift him or her over the threshold. In order to substantiate these considerations, we regress the manipulated fifth sub-score on the first four sub-scores and a set of course dummies. Results are presented in Table A3 of Appendix A. We find that the first four score components predict the interview score as expected (Column 1 and 2). In Column 3 we also add all possible interaction terms created from the four sub-scores of the model, which slightly improve its predictive power. Assuming that candidates' commitment and motivation are not (perfectly) correlated with education, being poor or being disadvantaged, the residual of this regression should now contain some relevant information on the selection committees' assessments of the candidates' aptitudes. We, therefore, use the predicted residual of the model in Column 3 to create the fifth component in the reconstructed assignment score.

[29] The treatment effects presented below are qualitatively and quantitatively robust to a number of alternative specifications, including the use of a different range of index scores to drop individuals, or adding other baseline characteristics to the model.

Because the points in the first four components were originally distributed by factors of five, we divide all score components by five to smooth the score and to minimize the heaping at multiples of five found in the original score. The final reconstructed score, therefore, ranges from 0 to 20.

### D. Simulating the Assignment Threshold

The assignment indicator we need for our analysis is not only determined by the individual scores, but also by the course-specific threshold score, which is likely to be affected by the manipulation of providers as well.[30] To construct a valid instrument, we re-estimate the threshold scores for each course following the approach proposed by Miller et al. (2013).[31] The authors' proposed approach depends on finding the optimal assignment variable based on a simulation exercise that maximizes the number of compliers given the reconstructed individual scores in a given course. Specifically, for each course, we run a set of simplified first stages similar to the one in Equation (1). We subsequently alter the threshold score, used to create the assignment dummy variable, $T_i$, from the lowest to the highest possible value. We then keep the threshold rule out of all possible assignment thresholds based on the specification that yields the largest $R^2$ for the respective course. Based on this optimal threshold, we proceed with calculating a reconstructed relative score, which serves as our new forcing variable.

### E. Balancing Performance of the Reconstructed Score

Our empirical approach assumes that no individual characteristics (other than vocational training enrollment) that could influence the outcomes of interest vary discontinuously across the estimated eligibility thresholds. As a first cut, and to assess whether the reconstruction of the score improves the differential sorting around the cut-off, we provide graphical evidence with respect to the density of the new forcing variable in Figure 4 (right graph). The density plot is significantly improved and does not appear to be discontinuous around the cut-off. We also employ the same balancing tests as before, now using the reconstructed score variable. Table 3's last two columns report the results. The new score successfully removes the imbalances we previously detected in all outcome variables as well as in the demographic characteristics at the baseline. Consistent with our assumption, estimates are not generally distinguishable from zero, except for the variables of age and Dalit ethnicity. In Figures 5 and 6, we present additional graphical evidence that outcomes and demographic characteristics are continuous around the cut-off of the running variable at baseline, except age and primary education. Age, being Dalit, and primary education are (practically) time-invariant characteristics in our sample. We, therefore, address the remaining unbalancing by following an estimation strategy on differenced outcomes.

---

[30] This is the case because the assignment threshold automatically moves with the distribution of the individual scores in each course.

[31] The authors follow Chay et al. (2005).

[Figure 5 and 6 about here]

Additionally, we show evidence in Figures 5 and 7 that the subgroup indicators we use to estimate heterogeneous treatment effects (i.e., an applicant's gender and the trade of training) are continuous across the threshold. This continuity across the threshold confirms that assignment status is not correlated with interaction variables around the cut-off, which is an important condition necessary for the estimation of unbiased heterogeneous treatment effects in the regression discontinuity setup (Becker et al., 2013).

[Figure 7 about here]

## VI. Results

### A. Program Effects on Employment and Earnings

We now turn our attention to the impact on the combined 2010, 2011 and 2012 samples based on the specifications described in Section V. Table 4, Panel A shows the local average treatment effects (LATE) estimated from Equations (1) and (2) – i.e., the effects the program had considering those individuals who complied with their assignment status – for employment and earnings on differenced outcomes using the *original* score. We find relatively large estimates across all outcomes, including significant (conditional) employment effects of 22 to 61 percentage points, as well as average earning gains of more than 3,000 NRs. The F-statistics (which range from 32 to 67) and the highly significant coefficients of the assignment variable ($\beta_1$) presented in the same panel suggest a strong first stage. However, as pointed out earlier, the estimates are likely to be biased upwards because the possible manipulation of the original score variable led to unbalanced individual characteristics and outcome variables at baseline.

[Table 4 about here]

Therefore, in Panel B, we use the *reconstructed* score, which led to a significantly improved covariate balancing. The first stage is very strong with F-statistics ranging from 38 to 88 and coefficients of the assignment variable ($\beta_1$) that are statistically significant at the 1-percent level. Comparing the effect sizes of Panels A and B (first rows, respectively) reveals that manipulation resulted in a strong upward bias in almost all outcomes. Using the corrected score, we no longer find evidence of a statistically significant impact on the employment rate, measured by whether individuals self-report any income-generating activities in the past month (Column 1, Panel B). When we restrict attention to employment in all non-farm activities, we find a lower but still statistically significant increase (Column 2, Panel B). The rate of participation in non-farm income-generating activities increases by 31 percentage points (from a base of 29 percent) as compared to 52 percentage points from using the manipulated score (Column 2, Panel A). Converting the results in percentage change terms, we find that the program increased non-farm employment by almost 94 percent. These impacts are not only statistically significant, but they are also economically meaningful. Disentangling these impacts into wage- and self-employment activities suggests that the effect is strongly driven by self-employment activities. The program increased non-farm self-employment by 30 percentage points, whereas we do not detect statistically significant impacts on non-farm wage employment rates (Columns 3 and 4, Panel B). We also examine the trade-specific income generating activity (IGA) rate – the percent of individuals who find employment in the same trade as the training that they applied for – and we find impacts of 40 percentage points (Column 5, Panel B). The increase in the employment rate within the same occupational fields for which the individuals trained implies that the skills trainees acquired in the vocational training part of the program were useful in starting employment activities within their respective trades. The EF program also led to improvements in the underemployment rate (i.e., cases in which people are working fewer hours than they wish). Column 6 of Panel B shows that the EF-sponsored training courses increased hours worked in IGAs by 49 hours per month (i.e., a 71 percent increase).

When we examine program impacts on monthly earnings using the corrected score, we find that effect sizes are about two-thirds of the estimates in Panel A. The reported results still show large program impacts (Columns 7 to 10, Panel B). We measure earnings as an individual's total earnings in the past month, including income from all IGAs, but not including unearned income.[32] We observe a statistically significant increase in monthly earnings for the treatment group by 1,754 NRs (≈ 23 USD), from a baseline average of 1,260 NRs. Once we restrict earnings only to those individuals who engage in an income generating activity, we detect an increase in earnings by 2,025 NRs (≈ 27 USD), from a baseline average of 2,075 NRs. In percentage terms, the increases in earnings translate to 140 and 98 percent, respectively. With alternative measurements of earnings, we also detect large program impacts. To account for the highly skewed nature of earnings distributions, we examine for impacts on logged earnings and find very sizable increases. In a third approach, we consider the proportion of participants who earned a "decent living." The EF considers 3,000 NRs per month (≈ 40 USD) as "gainful employment" or "being productively employed." At baseline, only 18 percent of the sample was gainfully employed. The EF training program increased this rate by 31 percentage points.

In Panels A and B, we use differenced outcomes to address the (remaining) disparities in participant characteristics we observed at baseline. In addition, the results reported in the discussion above are based on individuals who complied with their assignment status and, as a result, may differ from individuals in the full sample. In Panel C and D, we examine how altering these two features—that of differenced outcomes and the existence of compliers—affect our results. When comparing the estimates on differenced outcomes and on outcomes in levels, (Panel B versus C) using the reconstructed score, the difference in effect sizes is minor across most results. The most important difference we detect is on the variable that captures conditional earnings (Column 10, Panel C), which increases by approximately one-third compared to the estimate from the differenced outcome in Panel B. While the assigned group is slightly younger on average, they are also less likely to be Dalit than the rejected group. Older individuals may have been employed in a particular IGA for a longer time or, in general, may have more work experience, which is likely to lead to a higher level of earnings. On the other hand, individuals of Dalit ethnicity face substantial discrimination in the labor market, which likely makes them earn less compared to non-Dalits. The reduced effect size when controlling for time-invariant factors suggests that ethnicity outweighs the differences in age.

[32] If an individual did not work for pay in the past month, his/her earnings are recorded as zero.

Turning to Panel D, we examine the intent-to-treat (ITT) effects, based on Equation (3) in Section V. We also provide graphical evidence of these effects in Figure 8.

[Figure 8 about here]

As expected, we see sizeable differences: all ITT coefficients are substantially lower as compared to the LATE estimates. Although the ITT effects are smaller in size, they still indicate that training led to a significant increase in non-farm self-employment and trade-specific employment of 9 and 13 percentage points, an average increase of 15 hours per month of working time as well as an average rise in overall earnings of 572 NRs. The smaller effect sizes are to some extent due to the partial crossover between assignment groups (as we document above). Therefore, these coefficients can be interpreted as lower-bound estimates of the program effects.

Moreover, compliers in our sample may differ from non-compliers due to Nepal's labor market context, along with the program-specific targeting of particularly disadvantaged groups. Although the eligibility criteria of the program automatically exclude better-educated and non-poor individuals, the sample of applicants is still quite diverse along various characteristics that might affect the magnitude of the return an applicant can expect from participation in the program: in caste, in gender, in educational attainment, and in baseline intensity of poverty. It is entirely possible that the complier-population may be a subgroup of individuals for whom the returns to participation are larger than for the overall sample. For example, individuals who comply with program assignment may be otherwise unable to secure a self-paid seat in a similar program due to extreme poverty or caste- and gender-based discrimination. At the same time, these particular groups of applicants likely also have lower baseline educational levels and higher returns to additional education as compared to the rest of the sample. Therefore, we interpret the complier-based LATE coefficient estimates as upper-bound estimates of the program's effects.

## B. Gender-Disaggregated Program Effects

We also examine program impacts disaggregated for men and women (shown in Tables 5 and 6). The results reveal striking differences in the program's effectiveness by gender. For all outcomes, except for the conditional and unconditional earnings variables, the LATE estimates are larger and significantly different for women as compared to men. Except for the earnings conditional on employment outcome, none of the effects for males are statistically significant. The coefficients are, in fact, negative for most of the outcomes. In contrast, most effects for females are statistically significant, except for non-farm wage-employment, and earnings conditional on employment. The ITT results paint a similar picture and, additionally, show weak evidence of non-farm wage-employment gains for women. In summary, women seem to gain on the extensive margin across most outcomes (i.e., employment rates, hours worked, and earnings) without significant gains on the intensive margin (i.e*.,* conditional earnings). In comparison, men exhibit gains exclusively on the intensive margin.

[Table 5 about here]

Several factors could explain the differential gender impacts on employment outcomes. First, the EF introduced life-skills training for women, in 2010, in all of its training courses. Female students overwhelmingly responded positively to the life-skills training, often claiming that it was one of their favorite parts of the course. The skills learned and the positive experience in this life-skills training may contribute to the increased employment impact for women, which is line with the advice from experts in vocational training from around the world who increasingly advocate for the inclusion of life-skills in technical training programs.[33] Because all women received life-skills training, we ultimately cannot disentangle the influence of this particular factor from other program elements. Second, men exhibit higher non-farm baseline employment than women (49 percent compared to 18 percent for women) and therefore it is easier for women to make larger gains on the extensive margin. Third, women face higher labor market barriers as compared to men. We investigate this third potential explanation in Table 6, where we examine time allocation decisions by men and women for paid versus unpaid work, inside and outside the household, more closely. The results in Table 6 present a compelling story: program impacts on employment seem to be strongly driven by women who start self-employment activities *inside* the house, whereas unpaid work inside the house, and activities outside the house, remain unaffected by the program. These results suggest that the program is particularly effective at placing women into income generating activities while they remain at home – in this way, they conform to Nepal's social norms that restrict female mobility and bind them to household responsibilities.

[33] Acevedo et al. (2017), Adhvaryu et al. (2016), Bandiera et al. (2017), Ibarraran et al. (2014), and Martinez (2011) show significant impacts of life-skills interventions on labor market or productivity outcomes; Groh et al. (2012) find significant impacts of life-skills training intervention on women outside Jordan's capital; Ashraf et al. (2017) find significant impacts of a life-skills intervention among adolescent girls on subsequent human capital investment decisions in Zambia. In the Dominican Republic, Acevedo et al. (2017) examine the effectiveness of a life-skills intervention for men and women and, in fact, find higher effect sizes of the intervention on women. Adhvaryu et al. (2016), Bandiera et al. (2017), AGI (2013), USAID (2015), and Katz (2008) provide arguments for inclusion of life-skills (either alone or bundled with vocational training) targeting women in developing countries.

[Table 6 about here]

### C. Trade-Wise Program Effects

We examine program effects by type of trade and we classify training courses into nine categories. The most common training categories are: Tailoring/Garment/Textile (e.g., *galaicha* weaving, garment fabrication, hand embroidery, tailoring and dressmaking), Construction/Mechanical/Automobile (e.g., arc welding, brick molding, furniture making, motor bike service), and Electrical/Electronics/Computer (e.g., electric wiring, computer hardware technician, and mobile phone repair). These are followed by trainings related to Food preparation/Hospitality skills, Beautician/Barber skills, and Handicraft/Incense making skills. Finally, a few events in our sample are related to Farming, Poultry, and Security Guard skills training. Table 7 shows the breakdown of courses by trade. Although the EF specifically tried to encourage female participation in non-traditionally female trades, most of the training courses tended to be heavily gender-segregated. For example, men tended to dominate electronics and construction courses, whereas the tailoring, handicraft, and beautician training sessions were almost exclusively comprised of women.

[Table 7 about here]

Table 8 examines program impacts for the six most common training categories. Due to the low number of courses and participants, we dropped the remaining three categories from the analysis. Table 8 shows that the impacts of the skills training program differed markedly by type of trade. Trainings in beautician and tailoring consistently show strong impacts on employment—graduates of these training programs are more likely to have any (non-farm) employment and are also more likely to be working within the trade for which they were trained. Both trades also show large impacts on monthly hours worked and some of the earnings indicators.

[Table 8 about here]

For the remaining four trades, we do not detect conclusive, and significant, positive impacts. The coefficients of the trainings related to food preparation and hospitality even seem to reflect some negative influence of training on labor market outcomes (although these coefficients are mostly insignificant). Overall, the results in Tables 8 reveal substantial heterogeneity in employment and earning outcomes across the various types of training. The positive, and significant, impacts previously discussed are driven almost entirely by two categories of trades: beautician training and tailoring, both of which are almost entirely occupied by female trainees, which is in line with the large gender-differences in program impacts presented earlier.

## VII. Robustness Checks

Next, we present a number of consistency checks to investigate whether our estimates are within a reasonable range. We also provide additional robustness checks to address possible identification threats related to attrition issues or specification choices.

### A. Magnitude of Program Impacts

*Second Follow-up.* Most of the employment results we find are driven by self-employment activities and, hence, they do not seem to be a direct effect of the job-search-assistance component of the program. Still, providers could eagerly support trainees even in their entrepreneurial efforts in order to secure the last two installments of the EF payments. The EF verified placement 3 and 6 months upon training completion and released the final payment installment to providers, if and only if, they verified that trainees were in fact engaged in gainful employment. Therefore, it is possible that the employment effects we detect 9 to 11 months upon completion of training might be a mechanical effect of the program design. To examine this hypothesis, we investigate data on a sub-sample of training applicants, for whom a second follow-up survey (21 to 23 months after end of the training) was collected. Approximately half of the applicants from the 2010 cohort were randomly selected to participate in this second follow-up survey. The survey team tracked and re-interviewed 634 individuals (79.5 percent) who were also interviewed at baseline. Using data from the follow up survey, we re-run the specifications from Table 4, Panel B and D, as well as the ones presented Tables 5 and 6 on medium-term outcomes. Tables 9 and 10 present the results.

[Table 9 about here]

We find that most of the main effect estimates, presented in Table 9's Panel A and B, are in the range of the estimates we found using the first follow-up data for the pooled sample. Strikingly, some of the estimates are slightly larger. In particular, we find statistically significant estimates for non-farm employment, trade-specific employment, and gainful employment rates of 40, 33, and 40 percentage points, respectively, for the compliers (LATE) as well as 16, 14, and 16 percentage points, respectively, for the individuals assigned to training based on the intent-to-treat approach (ITT). Because the 2010 data do not include information on whether non-farm employment is based on wage- or self-employment, we cannot make inferences on what drives the results in that dimension. However, when we investigate the medium-term disaggregated effects for the 2010 cohort, we find a slightly different pattern in comparison to the pattern based on the short-term impacts for the pooled sample. Although the coefficients for females are still positive, slightly lower in magnitude, and insignificant (possibly a result of reduced statistical power), we now find positive employment effects for men. Specifically, the non-farm and trade-specific employment rates for male compliers rise by 50 and 48 percentage points, respectively, while the medium-term male ITT effects show an increase of 19 and 18 percentage points, respectively.

[Table 10 about here]

Program impacts on time-use (reported in Table 10) show negative but insignificant coefficient estimates for female paid work outside the house and positive but insignificant coefficients for women who perform paid work inside the house. For men, the coefficient estimates of both paid work inside and outside the house are positive but statistically insignificant. Although we do not detect statistically significant results, one could cautiously interpret them as somewhat suggestive evidence that medium-run employment rates for women, and to some extent for men, are also driven by self-employment activities.

*EF Verification Data.* For verification and monitoring purposes of program outcomes, the EF collected information on trainees' labor market outcomes. We retrieved information on these outcomes in aggregated form from four reports published by the EF from 2011 to 2013.[34] As another robustness check of our results, we compare these aggregated employment and earnings data from the EF with the data from our surveys (reported in Table 11).

[Table 11 about here]

Based on the aggregated data, the share of trainees employed was 91 percent, six months after training completion, for the 2012 graduates. The share of trainees in gainful employment (i.e., monthly income > 3,000 NRs) three and six months after completion of training was between 79 and 81 percent for the three cohorts. The employment rates, on the other hand, based on the self-reported data from our survey, range from 57 to 62 percent (9 to 11 months and 21 to 23 months after training completion). This comparison suggests that the numbers, based on the self-reported survey, are generally lower than those reported in the EF-verified data, which is consistent with the fact that providers receive their full payment after the six-month verification and that they have no additional incentives to support the job search of trainees. The rates for gainful employment are still relatively high among successful trainees at the time of our first and second follow-ups. Given that the employment rates at the three- and six-month mark are much higher, our estimates seem within a reasonable range.

In Figure 9, we also compare the earnings numbers based on the EF-verified data with the information from our first and second follow-up surveys. Given the growth trajectory of gainfully employed trainees' earnings over the first six months, post-training, the data based on the self-reported surveys seems reasonable and consistent with the overall pattern of the growth in earnings. Based both on this robustness check and also on the results from the 2010 cohort's second follow-up, we provide additional evidence consistent with the notion that our estimates are plausible.

[34] Data is available from the Employment Fund Annual Reports 2010 to 2012 as well as a tracer study of 2011 EF graduates, retrieved from http://www.employmentfund.org.np/category/resources/reports/ on Feb 24, 2018.

[Figure 9 about here]

## B. Differential Attrition

As in all panel studies, our estimates could be affected by a group-wise differential attrition. For instance, if trainees who are not admitted to the course choose to migrate for work abroad and if these are the most able and employable candidates of the control group, we would have an upward bias in our estimates. Therefore, in Table A4 of Appendix A, we explore the possibility of "*differential attrition*" and show no evidence to support it. Table A4 shows the results of a panel-based regression with attrition status as a dependent variable on a set of covariates for both the first follow-up (Columns 1 to 4) and the second-follow-up (Columns 5 to 8). All models are estimated based on the reconstructed score. In order to avoid the dropping of observations due to missing values in the control variables, Columns (3), (4), (7), and (8) include indicators that respectively flag missing values for each of the control variables. The regression results indicate that attrition is not correlated with assignment status in either wave.

## C. Alternative Bandwidths

To investigate the robustness of our results relative to bandwidth choice, we re-estimate our main specifications using bandwidths within 2, 3, 4, 5, and 10 index scores of the threshold based on the reconstructed score. Tables A5 presents the results, showing that overall our estimates are relatively stable both in statistical significance and in coefficient magnitude. For most outcomes, the magnitudes of the effect sizes increase with lower bandwidth choice. Only for non-farm self-employment are the effect sizes reduced by a half in the full sample. However, the effects for the female sub-sample are still large, positive, and significant using a bandwidth of 3 (ITT: 11, LATE: 36 percentage points, not shown in table) and, although still similar in size, they become insignificant at a bandwidth of 2 (ITT: 9, LATE: 32 percentage points, not shown in table).

## D. Propensity Score Approach

As a robustness check for estimates of program impact, we also employ a combination of differences-in-differences and a propensity score matching technique (Meyer, 1995). This approach, in the context of training programs, has the potential to purge possible differences between observable characteristics for trainees and non-trainees following Dehejia and Wahba (2002). The results based on this approach are presented in Appendix B, and they essentially confirm the pattern of the analysis using the regression discontinuity setup in the main paper.

## VIII. Discussion and Conclusion

Training interventions have been hailed as one potential solution to facilitate youth transition to productive employment and higher earnings worldwide. Although previous evaluations of training programs, based on observational designs, typically show positive and statistically significant impacts of training on the probability of having a job and on labor market earnings, recent experimental interventions from middle- and high-income countries find little or no impact on employment and modest gains in earnings. Using a regression discontinuity design in the context of a large vocational training program in Nepal, we find very large positive and statistically significant effects from the training program on female employment, hours worked, and earnings. These effects, in particular, are driven by women who engage in non-farm self-employment activities carried out inside (but not outside) the house.

In line with the few other existing studies on similar programs in low-income countries, our estimates of the employment effects of this training intervention are among the largest for training programs around the world. Features of the low-income background, the South Asian context, and the specific training intervention likely account for the large program impacts that we find. First, our program impacts are likely driven by a lack of alternative employment, skill training opportunities, and by extremely low education levels. Both of these phenomena are much more pronounced in the context of Nepal when compared with the context of the most recent experimental interventions from middle-income countries (i.e., in Latin America), especially so for women.

A second explanation behind the large program impacts relates to the extremely restrictive social norms regarding female labor force participation in Nepal. We find large program impacts on self-employment, especially inside the home, and these program effects are likely driven by social norms that prevent women from otherwise being active in the labor market. Restrictive norms regarding women entering the labor force exist in many regions of the world, but they seem to be particularly strong and restrictive for women in South Asia (World Bank, 2011; Asian Development Bank, 2015; Booth, 2016). Largely influenced by the Hindu philosophy on women's positions in society, Nepalese women face gender-based employment barriers due to restricted mobility, cultural norms, and societal norms as a result of gender expectations for certain occupations.

Finally, the EF training program was designed around employment outcomes such that training providers had to complete market assessments to ensure future employability with respect to the individually assigned trades. The training program was also bundled along with services, such as job placement, life-skills training, and business training, all of which likely also contributed to its effectiveness. Perhaps, most importantly, the program was directly tailored to the needs of the target population, and training providers were incentivized and closely monitored to accomplish their output.

Our results have important implications for the design and implementation of future training interventions in low-income countries. The empirical analysis presented here suggests important lessons for the successful modeling of effective labor market interventions where youth and female unemployment is a challenge.

# Figures

**FIGURE 1: TIMELINE**

| | 2010 | | 2011 | | 2012 | | 2013 |
|---|---|---|---|---|---|---|---|
| | **Jan-Apr** | | **Jan-Apr** | | **Jan-Apr** | | **Jan-Apr** |
| **2010 Cohort** | Baseline Survey /Start of Training | 1st EF Verif. 2nd EF Verif. | 1st Follow-up Survey | | 2nd Follow-up Survey* | | |
| **2011 Cohort** | | | Baseline Survey /Start of Training | 1st EF Verif. 2nd EF Verif. | 1st Follow-up Survey | | |
| **2012 Cohort** | | | | | Baseline Survey /Start of Training | 1st EF Verif. 2nd EF Verif. | 1st Follow-up Survey |

*Notes*: Baseline interviews were usually conducted a few days or weeks before start of training. Follow-up interviews were conducted approximately 12 and 24 months after baseline interviews, respectively. Training duration lasts from 1 to 3 months. EF verifies employment status and earnings of applicants 3 and 6 months after end of training, i.e., approx. 8 to 6 and 5 to 3 months before the first follow-up survey. *Roughly half of the initial sample of the 2010 cohort was randomly selected for the second follow-up survey.

**FIGURE 2: DISTRICTS COVERED BY COHORT**

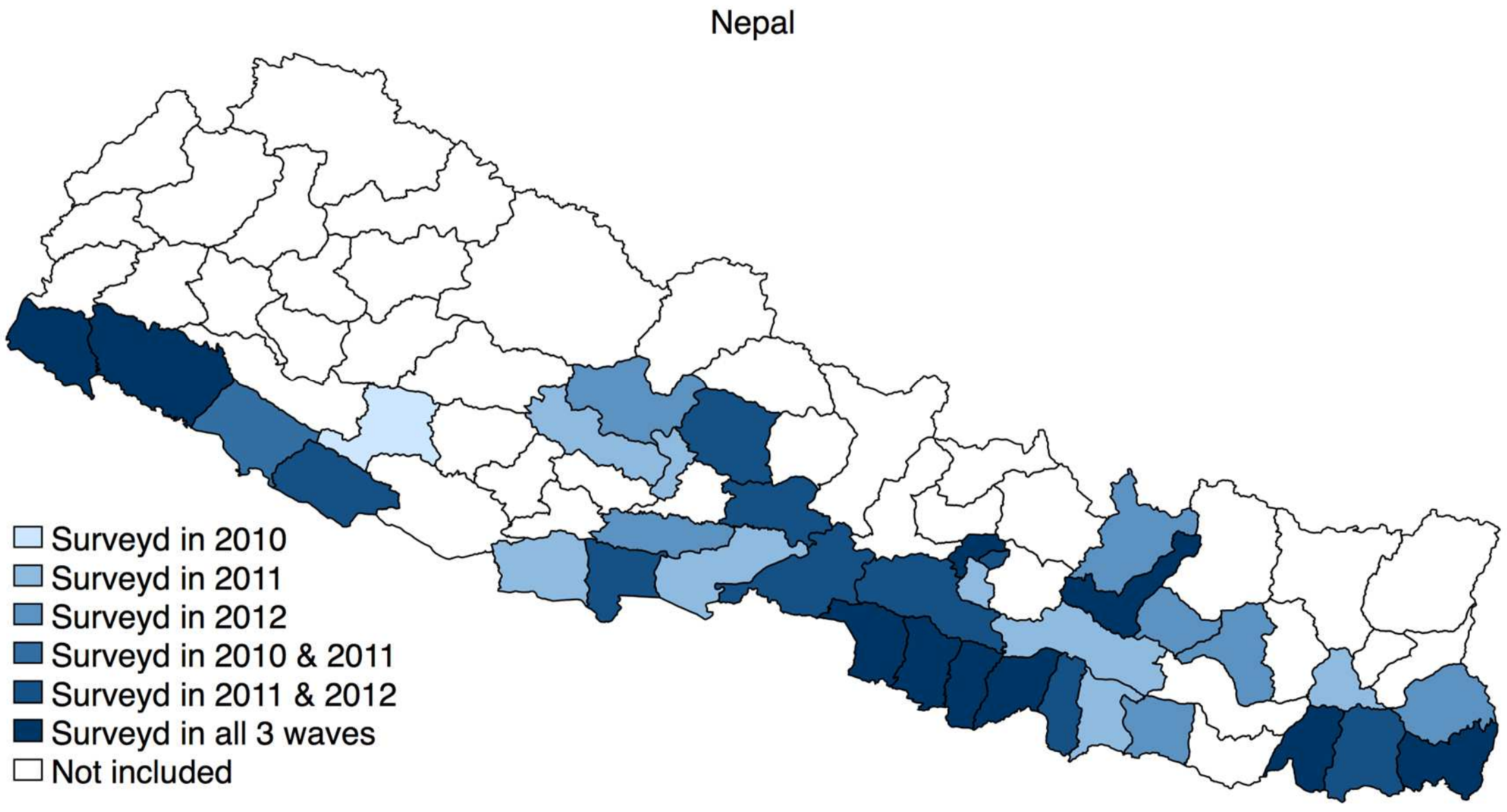

**FIGURE 3: PROBABILITY OF PROGRAM PARTICIPATION**

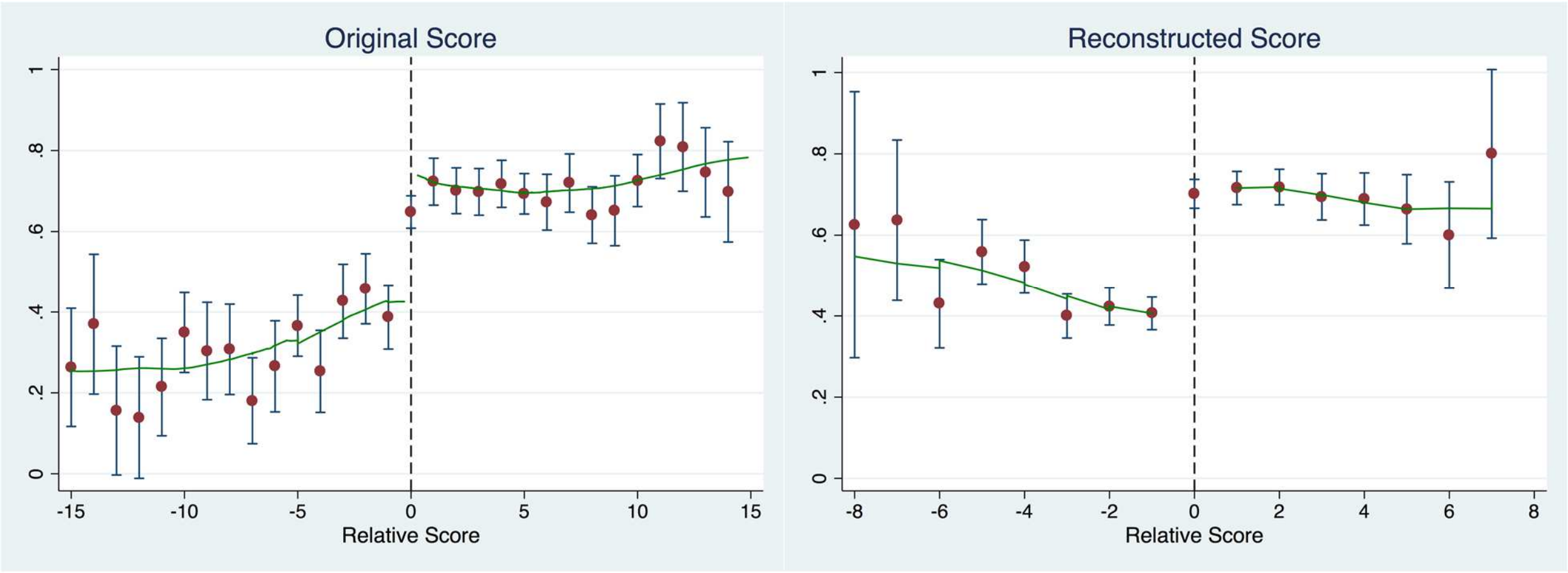

**FIGURE 4: DENSITY OF FORCING VARIABLE**

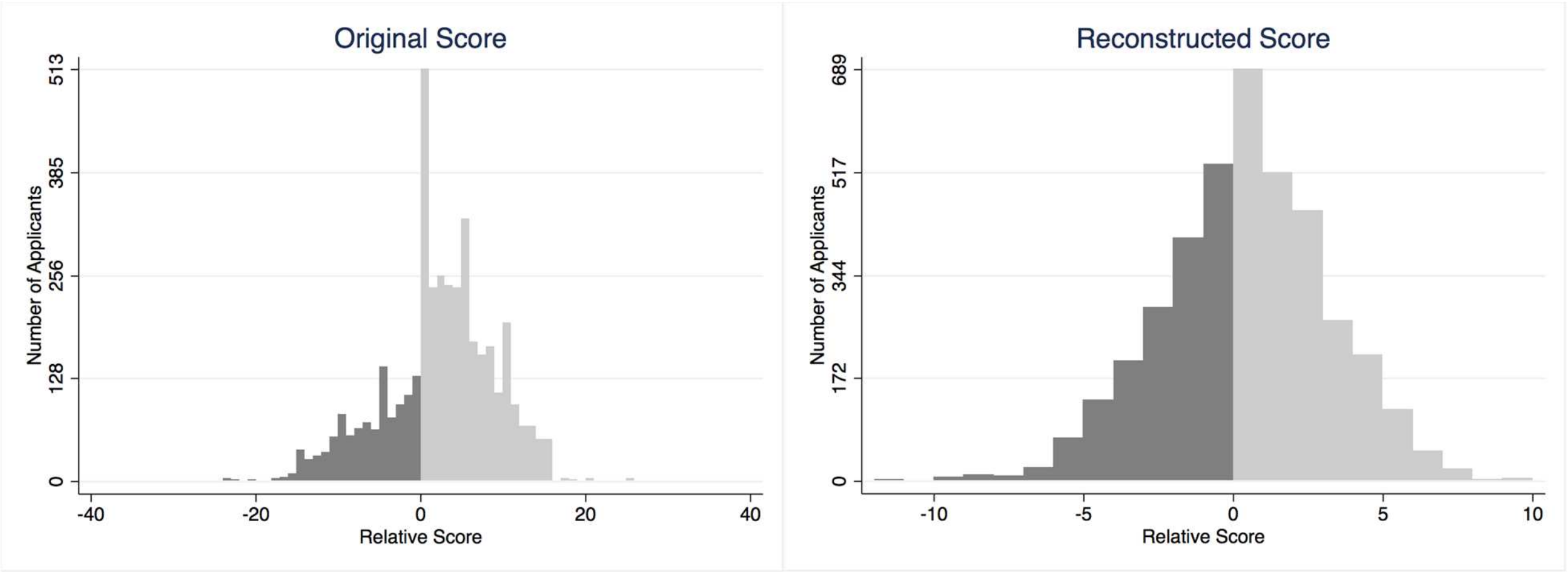

**FIGURE 5: CONTINUITY OF COVARIATES AROUND THE CUT-OFF AT BASELINE**

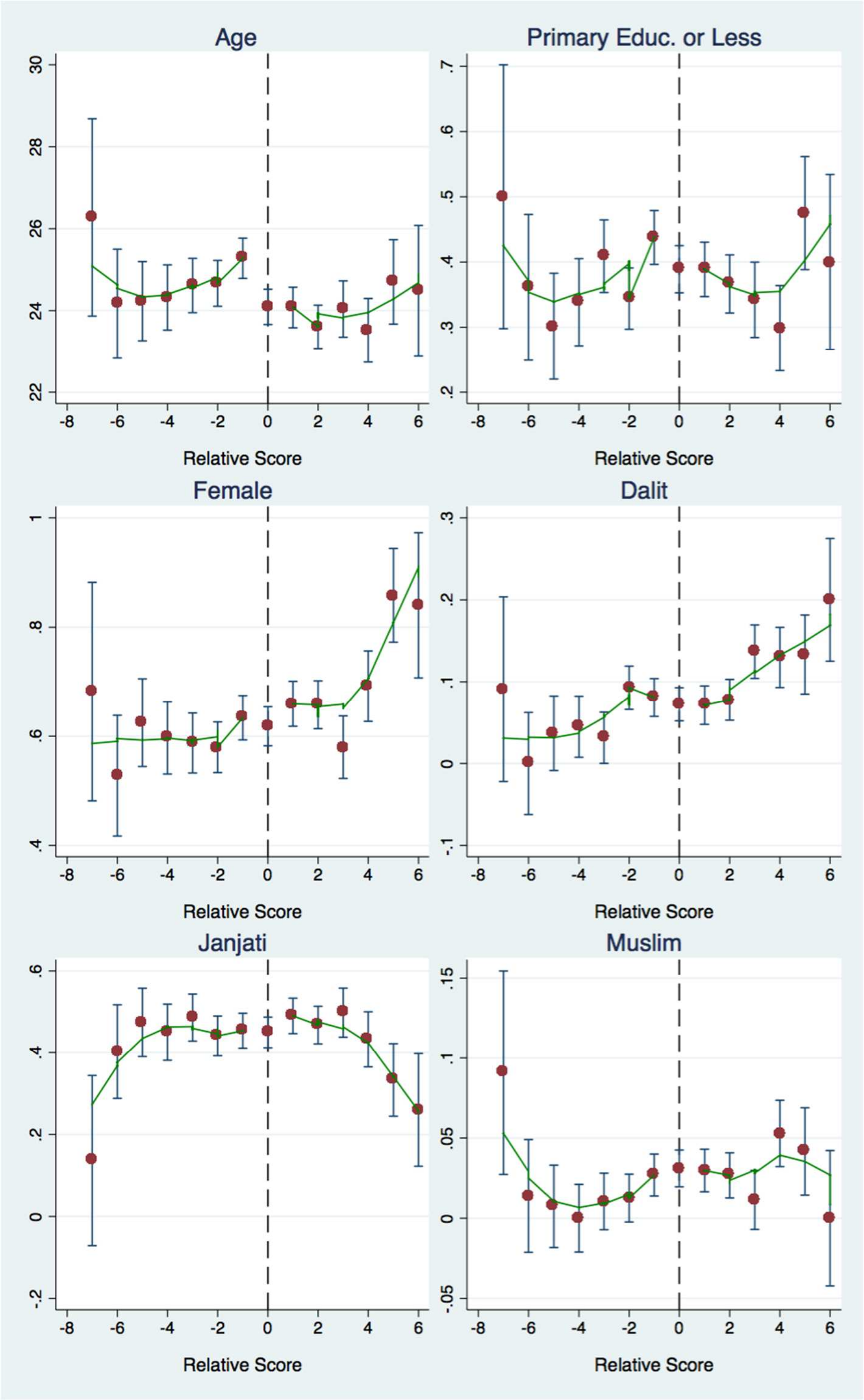

**FIGURE 6: CONTINUITY OF OUTCOMES AROUND THE CUT-OFF AT BASELINE**

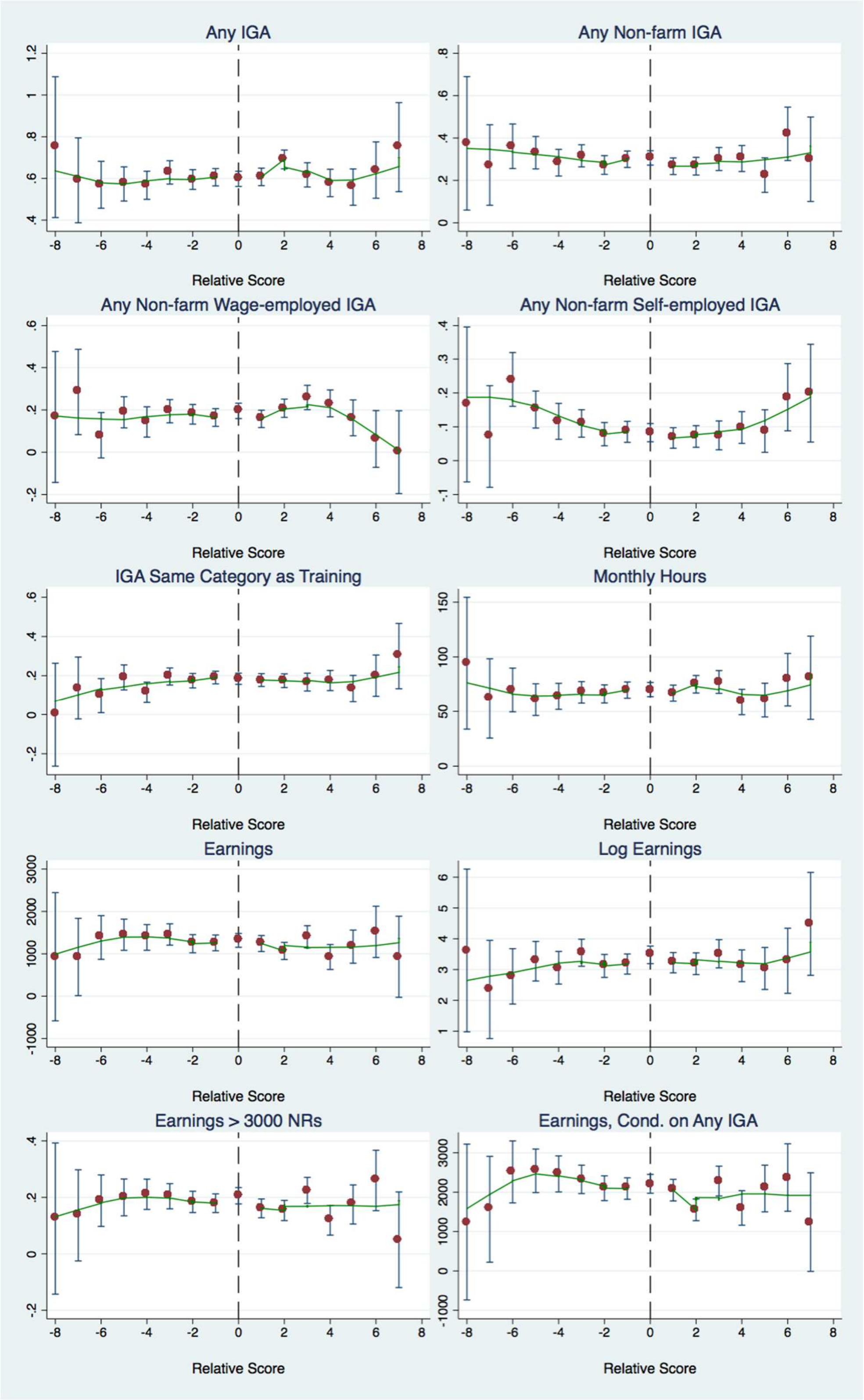

**FIGURE 7: CONTINUITY OF TRADE CHOICES AROUND THE CUT-OFF**

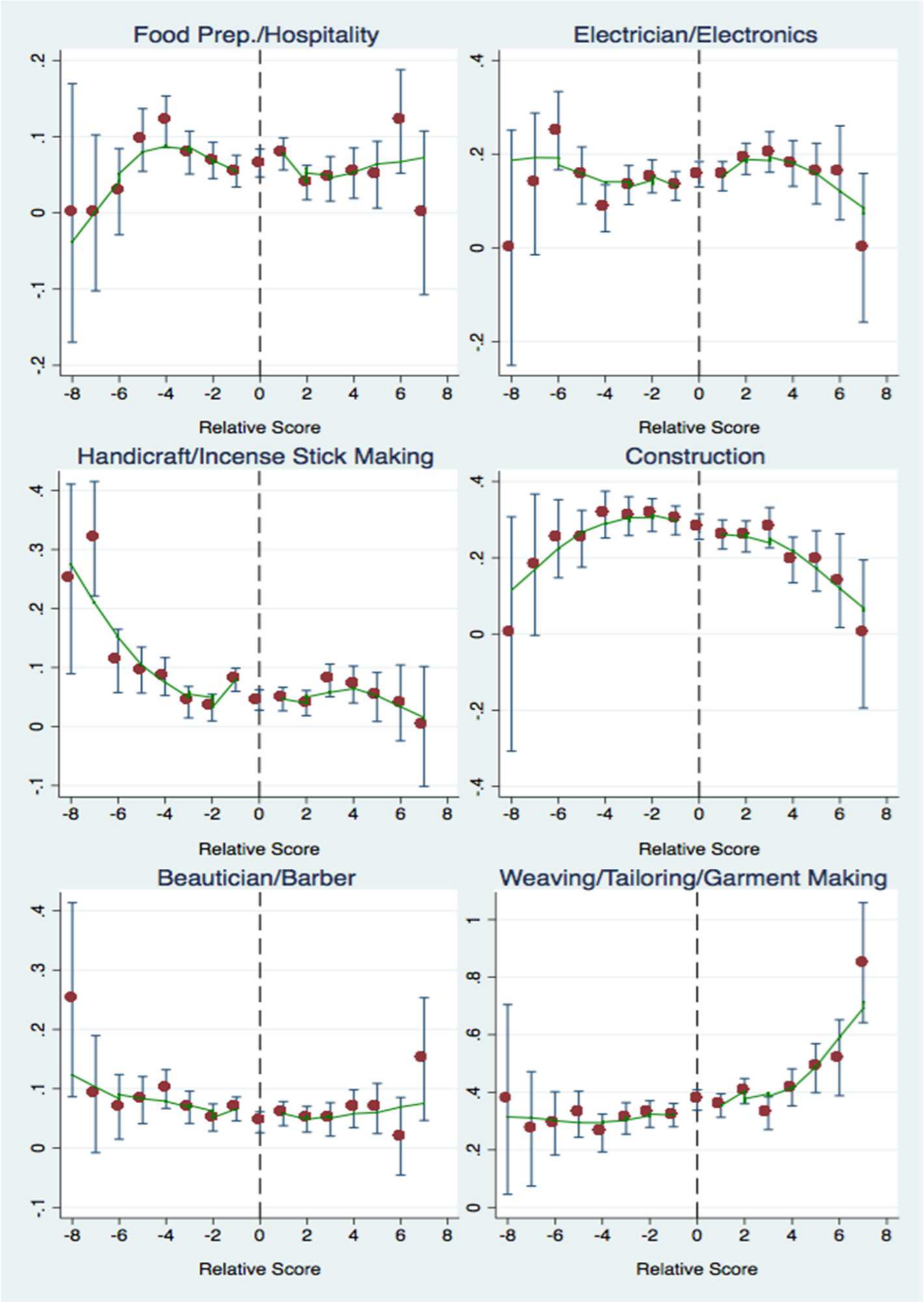

**FIGURE 8: GRAPHICAL REPRESENTATION OF INTENT-TO-TREAT RESULTS**

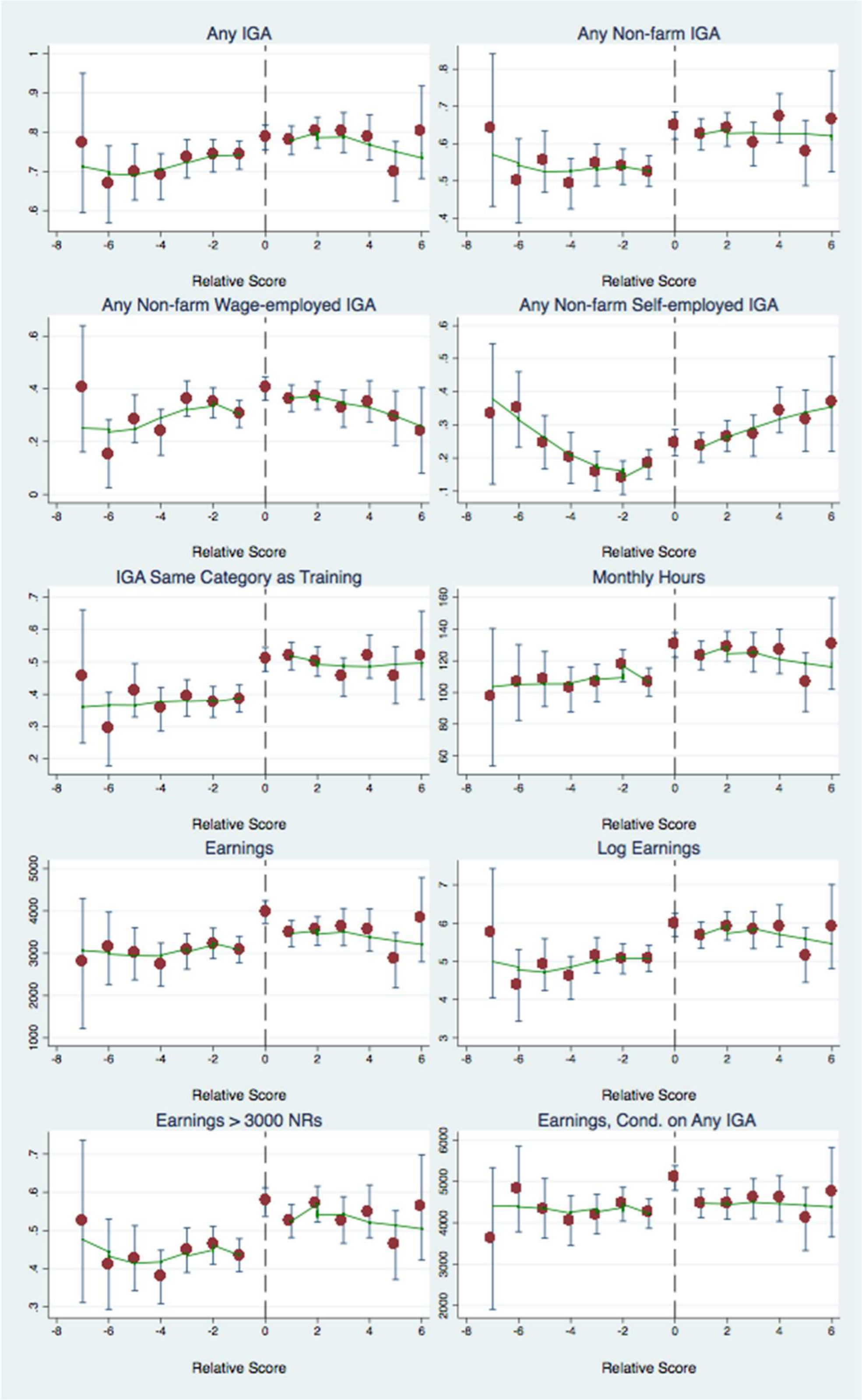

**FIGURE 9: EARNINGS BASED ON SURVEY AND ADMINISTRATIVE DATA COMPARISION**

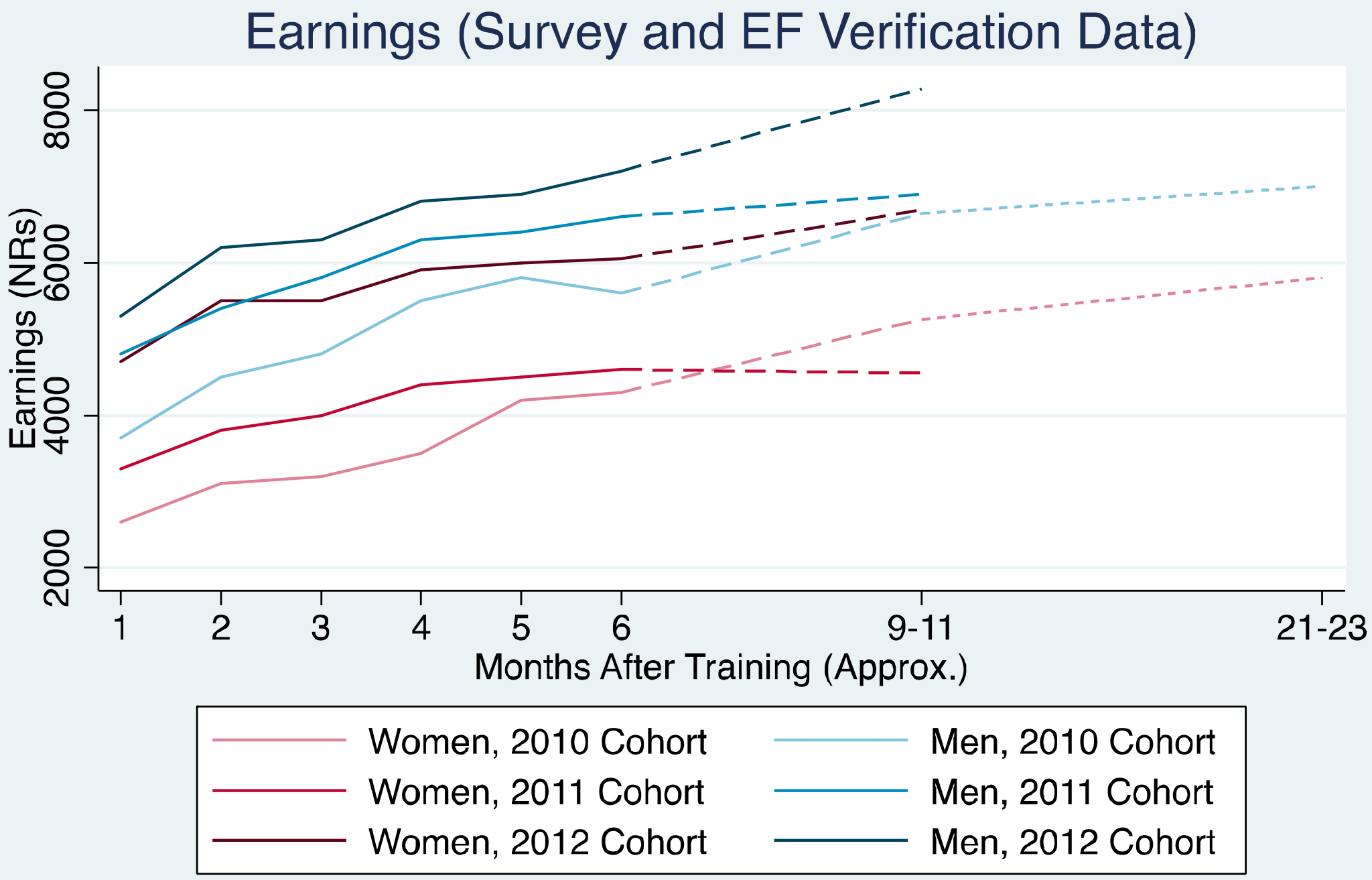

Note: Solid lines are based on EF verification data, retrieved from EF Annual Report 2012. Dashed lines are based on self-reported survey data approx. 9 to 11 and 21 to 23 month after training. Numbers represent average earnings of those trainees who are gainfully employed at a given month.

# Tables

**TABLE 1: TOTAL AND SAMPLED EF EVENTS**

| | 2010 | 2011 | 2012 | Pooled |
|---|---|---|---|---|
| **All EF events** | | | | |
| # Events conducted by EF (all year) | 598 | 645 | 711 | 1954 |
| # Events conducted by EF (Jan-Apr) | 110 | 142 | 143 | 395 |
| # Training providers | 21 | 32 | 35 | - |
| # Trainees | 11750 | 12869 | 14255 | 38874 |
| **Sampled EF events** | | | | |
| # Events randomly sampled for baseline survey | N/A | 182 | 112 | - |
| # Events included in baseline survey | 65 | 69 | 85 | 219 |
| # Districts covered | 12 | 28 | 26 | - |
| # Training providers covered | 18 | 26 | 28 | - |

*Notes*: More events were sampled than conducted in Jan-Apr 2011 because some events that were scheduled for Jan-Apr were delayed and did not start on time.

**TABLE 2: SURVEY RESPONSE RATES**

| | Baseline | Follow-up | Follow-up rate |
|---|---|---|---|
| **2010 cohort** | | | |
| Above Threshold | 1184 | 1047 | 88.4% |
| Below Threshold | 372 | 330 | 88.7% |
| **Total** | 1556 | 1377 | 88.5% |
| **2011 cohort** | | | |
| Above Threshold | 1237 | 1113 | 90.0% |
| Below Threshold | 349 | 306 | 87.7% |
| **Total** | 1586 | 1419 | 89.4% |
| **2012 cohort** | | | |
| Above Threshold | 1044 | 888 | 85.1% |
| Below Threshold | 491 | 417 | 84.9% |
| **Total** | 1535 | 1305 | 85.0% |
| **Pooled cohorts** | | | |
| Above Threshold | 3465 | 3049 | 88.0% |
| Below Threshold | 1212 | 1053 | 86.9% |
| **Total** | 4677 | 4101 | 87.7% |

*Notes*: The second follow-up survey for the 2010 cohort was conducted on a reduced (randomly selected) sample of 800 individuals. Out of those, the survey team was able to interview 634 individuals (79.5 percent) who were also interviewed at baseline.

**TABLE 3: BALANCE OF COVARIATES AND OUTCOME VARIABLES AT BASELINE (FULL SAMPLE)**

| Variable | Baseline Means | | | | | | | Differences (Original Score) | | | | Differences (Reconstructed Score) | | | |
|---|---|---|---|---|---|---|---|---|---|---|---|---|---|---|---|
| | All | Women | Men | Non-treated | Treated | Non-assigned | Assigned | LATE | | ITT | | LATE | | ITT | |
| **Demographics** | | | | | | | | | | | | | | | |
| Age | 24.32 | 24.66 | 23.72 | 24.39 | 24.27 | 24.34 | 24.31 | -1.21 | (1.53) | -0.32 | (0.42) | -4.37*** | (1.35) | -1.39*** | (0.41) |
| Primary Education or Less (1=Yes) | 0.38 | 0.43 | 0.29 | 0.39 | 0.37 | 0.35 | 0.39 | -0.16 | (0.12) | -0.04 | (0.03) | -0.12 | (0.09) | -0.04 | (0.03) |
| Men (1=Yes) | 0.37 | 0 | 1 | 0.37 | 0.37 | 0.36 | 0.37 | 0.38*** | (0.14) | 0.10*** | (0.04) | 0.00 | (0.10) | 0.00 | (0.03) |
| Dalit (1=Yes) | 0.08 | 0.08 | 0.08 | 0.10 | 0.07 | 0.09 | 0.08 | -0.38*** | (0.08) | -0.10*** | (0.02) | -0.11** | (0.06) | -0.03* | (0.02) |
| Janjati (1=Yes) | 0.46 | 0.42 | 0.51 | 0.41 | 0.49 | 0.42 | 0.47 | 0.09 | (0.14) | 0.02 | (0.04) | 0.04 | (0.11) | 0.02 | (0.03) |
| Muslim (1=Yes) | 0.03 | 0.03 | 0.02 | 0.02 | 0.03 | 0.02 | 0.03 | 0.01 | (0.04) | 0.00 | (0.01) | 0.04 | (0.04) | 0.01 | (0.01) |
| **Outcome Variables** | | | | | | | | | | | | | | | |
| Any IGA (1=Yes) | 0.61 | 0.52 | 0.77 | 0.59 | 0.63 | 0.59 | 0.62 | 0.11 | (0.12) | 0.03 | (0.03) | -0.04 | (0.10) | -0.01 | (0.03) |
| Any non-farm IGA (1=Yes) | 0.30 | 0.19 | 0.47 | 0.28 | 0.31 | 0.27 | 0.31 | 0.20* | (0.10) | 0.05* | (0.03) | -0.03 | (0.09) | -0.01 | (0.03) |
| Non-farm wage IGA (1=Yes) | 0.19 | 0.08 | 0.39 | 0.18 | 0.20 | 0.16 | 0.20 | 0.25** | (0.12) | 0.07** | (0.03) | 0.07 | (0.09) | 0.02 | (0.03) |
| Non-farm self IGA (1=Yes) | 0.09 | 0.10 | 0.09 | 0.09 | 0.09 | 0.09 | 0.10 | 0.07 | (0.09) | 0.02 | (0.02) | -0.02 | (0.07) | -0.01 | (0.02) |
| Trade-specific IGA (1=Yes) | 0.18 | 0.11 | 0.29 | 0.17 | 0.18 | 0.15 | 0.19 | 0.11 | (0.09) | 0.03 | (0.03) | -0.11 | (0.08) | -0.04 | (0.02) |
| Hours worked in past month | 69.26 | 47 | 108 | 66 | 71 | 63 | 72 | 40** | (19) | 11** | (5) | -1 | (18) | -0 | (6) |
| Earnings | 1269 | 770 | 2138 | 1254 | 1280 | 1180 | 1303 | 851* | (512) | 221 | (140) | -44 | (407) | -13 | (128) |
| Logged earnings | 3.29 | 2.43 | 4.80 | 3.15 | 3.39 | 3.03 | 3.39 | 1.37 | (0.96) | 0.36 | (0.26) | 0.15 | (0.73) | 0.05 | (0.23) |
| Earnings > 3000 NRs. (1=Yes) | 0.18 | 0.10 | 0.34 | 0.18 | 0.19 | 0.16 | 0.19 | 0.16* | (0.09) | 0.04* | (0.02) | 0.01 | (0.07) | 0.00 | (0.02) |
| Earnings, conditional on any IGA | 2082 | 1491 | 2774 | 2127 | 2054 | 2009 | 2109 | 943 | (735) | 256 | (209) | 201 | (608) | 57 | (190) |
| **Hours/week working ...** | | | | | | | | | | | | | | | |
| Unpaid, inside house > 5 (1=Yes) | 0.82 | 0.94 | 0.61 | 0.81 | 0.82 | 0.79 | 0.83 | 0.04 | (0.10) | 0.01 | (0.03) | -0.02 | (0.06) | -0.01 | (0.02) |
| Unpaid, inside house > 10 (1=Yes) | 0.63 | 0.82 | 0.31 | 0.62 | 0.64 | 0.60 | 0.65 | -0.00 | (0.13) | -0.00 | (0.04) | -0.10 | (0.09) | -0.03 | (0.03) |
| Unpaid, inside house > 20 (1=Yes) | 0.39 | 0.55 | 0.12 | 0.39 | 0.40 | 0.38 | 0.4 | -0.13 | (0.11) | -0.03 | (0.03) | -0.09 | (0.09) | -0.03 | (0.03) |
| Unpaid, outside house > 0 (1=Yes) | 0.61 | 0.56 | 0.69 | 0.61 | 0.61 | 0.60 | 0.61 | 0.18 | (0.12) | 0.05* | (0.03) | 0.07 | (0.09) | 0.02 | (0.03) |
| Paid, inside house > 0 (1=Yes) | 0.52 | 0.47 | 0.59 | 0.52 | 0.52 | 0.51 | 0.52 | 0.04 | (0.12) | 0.01 | (0.03) | -0.15 | (0.10) | -0.05 | (0.03) |
| Paid, outside house > 0 (1=Yes) | 0.48 | 0.36 | 0.69 | 0.45 | 0.49 | 0.45 | 0.49 | 0.04 | (0.12) | 0.01 | (0.03) | 0.00 | (0.10) | 0.00 | (0.03) |

*Notes*: *Non-farm wage IGA* and *Non-farm self IGA* are only available for 2011 and 2012 cohorts.

**TABLE 4: EMPLOYMENT**

| | Any IGA (1=Yes) | Any non-farm IGA (1=Yes) | Non-farm wage IGA (1=Yes) | Non-farm self IGA (1=Yes) | Trade-specific IGA (1=Yes) | Hours worked in past month | Earnings | Logged earnings | Earnings > 3000 NRs. (1=Yes) | Earnings, conditional on any IGA |
|---|---|---|---|---|---|---|---|---|---|---|
| | (1) | (2) | (3) | (4) | (5) | (6) | (7) | (8) | (9) | (10) |
| **Panel A** | | | | | | | | | | |
| LATE, DD | 0.27* | 0.52*** | 0.38*** | 0.22* | 0.61*** | 75.92*** | 3074*** | 2.85* | 0.43*** | 3179** |
| (Original Score) | (0.15) | (0.15) | (0.15) | (0.12) | (0.14) | (25.57) | (987) | (1.55) | (0.14) | (1295) |
| First Stage | 0.27*** | 0.26*** | 0.29*** | 0.29*** | 0.27*** | 0.27*** | 0.26*** | 0.23*** | 0.26*** | 0.28*** |
| | (0.03) | (0.04) | (0.04) | (0.04) | (0.03) | (0.03) | (0.04) | (0.04) | (0.04) | (0.05) |
| F-statistic | 67.30 | 54.90 | 47.53 | 47.53 | 58.42 | 67.30 | 53.73 | 36.04 | 52.31 | 32.37 |
| Baseline mean | 0.61 | 0.30 | 0.19 | 0.09 | 0.18 | 69.26 | 1271 | 3.27 | 0.18 | 2084 |
| Bandwidth | 37 | 15 | 37 | 37 | 19 | 37 | 23 | 12 | 19 | 26 |
| Observations | 4080 | 3976 | 2677 | 2677 | 4066 | 4080 | 3899 | 3544 | 3891 | 1979 |
| **Panel B** | | | | | | | | | | |
| LATE, DD | 0.13 | 0.31*** | 0.11 | 0.30** | 0.40*** | 48.85** | 1754** | 1.75** | 0.31*** | 2025* |
| (Reconstr. Sc.) | (0.10) | (0.10) | (0.11) | (0.12) | (0.12) | (20.50) | (696) | (0.87) | (0.11) | (1115) |
| First Stage | 0.32*** | 0.32*** | 0.31*** | 0.30*** | 0.32*** | 0.32*** | 0.33*** | 0.32*** | 0.32*** | 0.30*** |
| | (0.03) | (0.03) | (0.04) | (0.04) | (0.03) | (0.03) | (0.04) | (0.03) | (0.03) | (0.05) |
| F-statistic | 87.47 | 87.47 | 54.84 | 51.12 | 87.47 | 87.47 | 85.92 | 86.33 | 86.33 | 38.23 |
| Baseline mean | 0.61 | 0.29 | 0.18 | 0.09 | 0.18 | 68.62 | 1260 | 3.27 | 0.18 | 2075 |
| Bandwidth | 12 | 12 | 12 | 11 | 12 | 12 | 9 | 12 | 12 | 12 |
| Observations | 4004 | 4004 | 2608 | 2606 | 4004 | 4004 | 3822 | 3838 | 3838 | 1931 |
| **Panel C** | | | | | | | | | | |
| LATE, Level | 0.09 | 0.33*** | 0.17 | 0.26** | 0.33*** | 48.08** | 1934*** | 1.80** | 0.32*** | 2686** |
| (Reconstr. Sc.) | (0.08) | (0.09) | (0.11) | (0.12) | (0.10) | (21.01) | (745) | (0.81) | (0.10) | (1355) |
| First Stage | 0.32*** | 0.34*** | 0.31*** | 0.31*** | 0.34*** | 0.32*** | 0.34*** | 0.32*** | 0.32*** | 0.29*** |
| | (0.03) | (0.04) | (0.04) | (0.05) | (0.04) | (0.03) | (0.04) | (0.03) | (0.03) | (0.05) |
| F-statistic | 87.47 | 88.14 | 57.61 | 35.82 | 88.14 | 87.47 | 89.99 | 88.26 | 88.26 | 29.78 |
| Baseline mean | 0.61 | 0.29 | 0.18 | 0.08 | 0.18 | 68.62 | 1261 | 3.27 | 0.18 | 2043 |
| Bandwidth | 12 | 8 | 12 | 4 | 8 | 12 | 8 | 12 | 12 | 3 |
| Observations | 4004 | 3976 | 2633 | 2164 | 3976 | 4004 | 3836 | 3864 | 3864 | 2110 |
| **Panel D** | | | | | | | | | | |
| ITT, DD | 0.04 | 0.10*** | 0.03 | 0.09** | 0.13*** | 15.41** | 572** | 0.57** | 0.10*** | 598* |
| (Reconst. Sc.) | (0.03) | (0.03) | (0.03) | (0.04) | (0.04) | (6.45) | (227) | (0.28) | (0.03) | (324) |
| **Group means at follow-up (Within given Bandwidth of Panel B)** | | | | | | | | | | |
| Non-treated | 0.69 | 0.46 | 0.28 | 0.16 | 0.30 | 98.36 | 2820 | 4.54 | 0.40 | 4157 |
| Treated | 0.81 | 0.68 | 0.38 | 0.29 | 0.55 | 132.79 | 3780 | 6.13 | 0.57 | 4717 |
| Non-assigned | 0.72 | 0.53 | 0.31 | 0.19 | 0.37 | 107.78 | 3035 | 4.96 | 0.43 | 4262 |
| Assigned | 0.79 | 0.64 | 0.36 | 0.27 | 0.50 | 126.56 | 3634 | 5.84 | 0.55 | 4665 |

*Notes*: Standard errors adjusted for clustering at event level reported in parentheses. Each cell represents an estimate from a separate regression, which includes the total score, the relative score (forcing variable), and an interaction of the relative score with the assignment variable as counterfactuals, as well as a constant. For comparability, Panel D estimates and follow-up means are conducted within the same bandwidths as in Panel B. Assigned/Non-assigned group means are based on the reconstructed score. *Non-farm wage IGA* and *Non-farm self IGA* are only available for 2011 and 2012 cohorts. ***, **, and * denote significance at the 1, 5, and 10 percent level.

**TABLE 5: EMPLOYMENT BY GENDER**

| | Any IGA (1=Yes) | Any non-farm IGA (1=Yes) | Non-farm wage IGA (1=Yes) | Non-farm self IGA (1=Yes) | Trade-specific IGA (1=Yes) | Hours worked in past month | Earnings | Logged earnings | Earnings > 3000 NRs. (1=Yes) | Earnings, conditional on any IGA |
|---|---|---|---|---|---|---|---|---|---|---|
| | (1) | (2) | (3) | (4) | (5) | (6) | (7) | (8) | (9) | (10) |
| **Panel A**: HLATE, DD (Reconstructed Score) | | | | | | | | | | |
| Women | 0.26* | 0.53*** | 0.21 | 0.46*** | 0.54*** | 90.43*** | 2113*** | 3.51*** | 0.52*** | 1590 |
| | (0.14) | (0.14) | (0.13) | (0.15) | (0.15) | (24.67) | (784) | (1.11) | (0.13) | (1256) |
| Men | -0.09 | -0.04 | -0.11 | 0.03 | 0.18 | -18.63 | 977 | -1.20 | -0.04 | 2461* |
| | (0.12) | (0.15) | (0.17) | (0.13) | (0.16) | (27.33) | (1028) | (1.10) | (0.14) | (1470) |
| Difference | 0.35** | 0.56*** | 0.32* | 0.43** | 0.36* | 109.1*** | 1136 | 4.71*** | 0.56*** | -871 |
| | (0.16) | (0.19) | (0.18) | (0.17) | (0.20) | (30.65) | (1143) | (1.40) | (0.17) | (1560) |
| Bandwidth | 12 | 12 | 12 | 11 | 12 | 12 | 9 | 12 | 12 | 12 |
| Observations | 4004 | 4004 | 2608 | 2606 | 4004 | 4004 | 3822 | 3838 | 3838 | 1931 |
| **Group means at Baseline (Within a given Bandwidth)** | | | | | | | | | | |
| Women | 0.52 | 0.19 | 0.07 | 0.10 | 0.11 | 46.94 | 770 | 2.43 | 0.10 | 1492 |
| Men | 0.77 | 0.47 | 0.38 | 0.09 | 0.29 | 106.90 | 2130 | 4.77 | 0.34 | 2778 |
| **Panel B**: Heterogeneous ITT, DD (Reconstructed Score) | | | | | | | | | | |
| Women | 0.08* | 0.16*** | 0.07* | 0.14*** | 0.16*** | 27.15*** | 659*** | 1.06*** | 0.16*** | 488 |
| | (0.04) | (0.04) | (0.04) | (0.04) | (0.05) | (7.01) | (242) | (0.33) | (0.04) | (381) |
| Men | -0.01 | 0.01 | -0.02 | 0.02 | 0.07 | -1.88 | 386 | -0.20 | 0.01 | 698* |
| | (0.04) | (0.04) | (0.04) | (0.04) | (0.05) | (8.10) | (310) | (0.32) | (0.04) | (392) |
| Difference | 0.09** | 0.15*** | 0.09* | 0.12** | 0.09 | 29.03*** | 273 | 1.25*** | 0.15*** | -210 |
| | (0.04) | (0.05) | (0.05) | (0.05) | (0.06) | (8.01) | (313) | (0.37) | (0.04) | (420) |

*Notes*: Standard errors adjusted for clustering at event level reported in parentheses. Panel A presents second stage results. In both panels, each cell represents a treatment effect retrieved from a regression, which includes the group variable, an interaction term of the group variable with the training or assignment indicator, the total score, the relative score (forcing variable), and an interaction of the relative score with the assignment variable as counterfactuals, as well as a constant. *Difference* is the coefficient of the interaction term with the group variable from the respective regression. For comparability, all estimates are conducted within the same bandwidths as in Panel B, Table 4. *Non-farm wage IGA* and *Non-farm self IGA* are only available for 2011 and 2012 cohorts. ***, **, and * denote significance at the 1, 5, and 10 percent level.

**TABLE 6: TIME USE BY GENDER**

| | Hours/week working: Unpaid, inside house > 5 (1=Yes) | Unpaid, inside house > 10 (1=Yes) | Unpaid, inside house > 20 (1=Yes) | Unpaid, outside house > 0 (1=Yes) | Paid, inside house > 0 (1=Yes) | Paid, outside house > 0 (1=Yes) |
|---|---|---|---|---|---|---|
| | (1) | (2) | (3) | (4) | (5) | (6) |
| **Panel A**: HLATE, DD (Reconstructed Score) | | | | | | |
| Women | -0.13 | -0.12 | 0.10 | 0.03 | 0.42*** | 0.07 |
| | (0.09) | (0.12) | (0.22) | (0.16) | (0.15) | (0.12) |
| Men | 0.04 | 0.02 | 0.15 | -0.11 | 0.08 | -0.08 |
| | (0.16) | (0.15) | (0.20) | (0.17) | (0.17) | (0.13) |
| Difference | -0.17 | -0.14 | -0.06 | 0.15 | 0.34 | 0.16 |
| | (0.17) | (0.17) | (0.17) | (0.22) | (0.21) | (0.16) |
| Bandwidth | 12 | 12 | 2 | 12 | 12 | 12 |
| Observations | 4003 | 4003 | 2133 | 4000 | 4004 | 4002 |
| **Group means at Baseline (Within a Given Bandwidth)** | | | | | | |
| Women | 0.94 | 0.82 | 0.60 | 0.56 | 0.47 | 0.36 |
| Men | 0.61 | 0.32 | 0.11 | 0.70 | 0.59 | 0.69 |
| **Panel B**: Heterogeneous ITT, DD (Reconstructed Score) | | | | | | |
| Women | -0.04 | -0.04 | 0.03 | 0.00 | 0.13*** | 0.02 |
| | (0.03) | (0.03) | (0.06) | (0.05) | (0.04) | (0.03) |
| Men | 0.01 | -0.00 | 0.05 | -0.03 | 0.04 | -0.02 |
| | (0.05) | (0.05) | (0.06) | (0.05) | (0.05) | (0.04) |
| Difference | -0.05 | -0.04 | -0.02 | 0.03 | 0.09 | 0.04 |
| | (0.05) | (0.05) | (0.05) | (0.06) | (0.06) | (0.04) |

*Notes*: Standard errors adjusted for clustering at event level reported in parentheses. Panel A presents second stage results. In both panels, each cell represents a treatment effect retrieved from a regression, which includes the group variable, an interaction term of the group variable with the training or assignment indicator, the total score, the relative score (forcing variable), and an interaction of the relative score with the assignment variable as counterfactuals, as well as a constant. *Difference* is the coefficient of the interaction term with the group variable from the respective regression. ***, **, and * denote significance at the 1, 5, and 10 percent level.

## TABLE 7: TYPES OF TRAINING

| | By Event | | By Applicant | | Female Applicants |
|---|---|---|---|---|---|
| | Frequency | Percent | Frequency | Percent | Percent |
| Farming | 5 | 2 | 92 | 2 | 88 |
| Poultry | 2 | 1 | 41 | 1 | 100 |
| Food Prep/Hospitality | 16 | 7 | 260 | 6 | 54 |
| Electrical/Electronics/Computer | 37 | 17 | 639 | 16 | 18 |
| Handicraft & Incense | 12 | 5 | 233 | 6 | 89 |
| Construction/Mechanical/Automobile | 63 | 29 | 1127 | 27 | 30 |
| Beautician /Barber | 11 | 5 | 239 | 6 | 100 |
| Tailoring/Garment/Textile | 72 | 33 | 1457 | 36 | 98 |
| Security Guard | 1 | 0 | 13 | 0 | 100 |
| **Total** | **219** | **100** | **4101** | **100** | |

**TABLE 8: EMPLOYMENT BY TRADE**

| | Any IGA (1=Yes) | Any non-farm IGA (1=Yes) | Non-farm wage IGA (1=Yes) | Non-farm self IGA (1=Yes) | Trade-specific IGA (1=Yes) | Hours worked in past month | Earnings | Logged earnings | Earnings > 3000 NRs. (1=Yes) | Earnings, conditional on any IGA |
|---|---|---|---|---|---|---|---|---|---|---|
| | (1) | (2) | (3) | (4) | (5) | (6) | (7) | (8) | (9) | (10) |
| **Panel A**: HLATE, DD (Reconstructed Score) | | | | | | | | | | |
| Food prep. & Hospitality | -0.34 | -0.62 | -1.73 | 1.64*** | -0.73 | -101.23 | -5608 | -4.98 | -0.43 | -156 |
| | (0.43) | (0.64) | (1.14) | (0.55) | (0.69) | (110.95) | (4626) | (5.22) | (0.56) | (6068) |
| Electrician & Electronics | -0.25 | 0.04 | 0.02 | 0.03 | 0.42*** | -48.10 | -172 | -1.53 | -0.06 | 2028 |
| | (0.19) | (0.17) | (0.19) | (0.17) | (0.16) | (35.92) | (1280) | (1.39) | (0.18) | (1644) |
| Handicraft & Incense stick making | 0.11 | -0.58 | -0.22 | 0.25 | -1.02 | 51.40 | 1150 | -2.05 | 0.18 | 1715 |
| | (0.47) | (0.57) | (0.14) | (0.24) | (0.80) | (56.27) | (2666) | (3.55) | (0.30) | (2900) |
| Construction & Mechanics | -0.08 | -0.03 | -0.01 | -0.07 | 0.12 | 11.47 | 139 | -1.49 | 0.00 | 1716 |
| | (0.11) | (0.17) | (0.21) | (0.13) | (0.20) | (29.32) | (1139) | (1.27) | (0.16) | (1407) |
| Beautician & Barber | 1.08* | 1.06* | 0.67** | 0.34 | 0.99*** | 205.57** | 4379 | 6.63* | 0.56** | -1391 |
| | (0.65) | (0.54) | (0.29) | (0.29) | (0.37) | (89.18) | (2721) | (3.62) | (0.27) | (4295) |
| Weaving, Tailoring & Garment Making | 0.38** | 0.82*** | 0.30* | 0.57*** | 0.86*** | 112.5*** | 3572*** | 5.70*** | 0.75*** | 2985* |
| | (0.18) | (0.18) | (0.18) | (0.20) | (0.20) | (32.10) | (1127) | (1.59) | (0.19) | (1621) |
| **Panel B**: Heterogeneous ITT, DD (Reconstructed Score) | | | | | | | | | | |
| Food prep. & Hospitality | -0.05 | -0.09 | -0.31*** | 0.31 | -0.10 | -14.68 | -924* | -0.83 | -0.06 | -10 |
| | (0.08) | (0.10) | (0.09) | (0.20) | (0.09) | (18.02) | (545) | (0.82) | (0.09) | (836) |
| Electrician & Electronics | -0.08 | 0.02 | 0.01 | 0.01 | 0.15** | -13.52 | 28 | -0.42 | -0.01 | 639 |
| | (0.06) | (0.06) | (0.06) | (0.05) | (0.06) | (11.24) | (433) | (0.44) | (0.06) | (524) |
| Handicraft & Incense stick making | 0.04 | -0.12 | -0.09 | 0.10 | -0.22* | 16.38 | 424 | -0.34 | 0.07 | 675 |
| | (0.12) | (0.11) | (0.07) | (0.10) | (0.12) | (14.99) | (629) | (0.77) | (0.07) | (1018) |
| Construction & Mechanics | -0.02 | 0.01 | 0.00 | 0.00 | 0.06 | 6.56 | 526 | -0.29 | 0.02 | 552 |
| | (0.04) | (0.05) | (0.05) | (0.03) | (0.06) | (9.25) | (347) | (0.38) | (0.05) | (383) |
| Beautician & Barber | 0.25** | 0.26*** | 0.21*** | 0.11 | 0.25*** | 49.38*** | 1096* | 1.62** | 0.15** | -366 |
| | (0.11) | (0.08) | (0.07) | (0.09) | (0.07) | (18.88) | (600) | (0.77) | (0.06) | (1165) |
| Weaving, Tailoring & Garment Making | 0.11** | 0.24*** | 0.09* | 0.16*** | 0.25*** | 32.32*** | 1043*** | 1.64*** | 0.22*** | 985* |
| | (0.05) | (0.04) | (0.05) | (0.05) | (0.05) | (8.09) | (304) | (0.39) | (0.05) | (509) |

*Notes*: Standard errors adjusted for clustering at event level reported in parentheses. Panel A presents second stage results. In both panels, each cell represents an estimate from a separate regression, which includes the group variable, an interaction term of the group variable with the training or assignment indicator, the total score, the relative score (forcing variable), and an interaction of the relative score with the assignment variable as counterfactuals, as well as a constant. For comparability, all estimates are conducted within the same bandwidths as in Panel B, Table 4. *Non-farm wage IGA* and *Non-farm self IGA* are only available for 2011 and 2012 cohorts. ***, **, and * denote significance at the 1, 5, and 10 percent level.

**TABLE 9: EMPLOYMENT, SECOND FOLLOW-UP (2010 COHORT)**

| | Any IGA (1=Yes) | Any non-farm IGA (1=Yes) | Trade-specific IGA (1=Yes) | Hours worked in past month | Earnings | Logged earnings | Earnings > 3000 NRs. (1=Yes) | Earnings, conditional on any IGA |
|---|---|---|---|---|---|---|---|---|
| | (1) | (2) | (3) | (4) | (5) | (6) | (7) | (8) |
| **Panel A** | | | | | | | | |
| LATE, DD | 0.06 | 0.40** | 0.33** | 53.16 | 1277 | 0.65 | 0.40* | 903 |
| (Reconstr. Sc.) | (0.17) | (0.19) | (0.16) | (32.90) | (1362) | (1.52) | (0.23) | (1510) |
| First Stage | 0.41*** | 0.41*** | 0.41*** | 0.41*** | 0.41*** | 0.41*** | 0.41*** | 0.47*** |
| | (0.06) | (0.06) | (0.06) | (0.06) | (0.06) | (0.06) | (0.06) | (0.08) |
| F-statistic | 48.33 | 48.33 | 48.33 | 48.33 | 42.58 | 42.58 | 42.58 | 36.07 |
| Baseline mean | 0.61 | 0.31 | 0.18 | 81.37 | 1388 | 3.63 | 0.20 | 2304 |
| Bandwidth | 10 | 10 | 10 | 10 | 10 | 10 | 10 | 10 |
| Observations | 621 | 621 | 621 | 621 | 590 | 590 | 590 | 336 |
| **Panel B** | | | | | | | | |
| ITT, DD | 0.01 | 0.16* | 0.14** | 20.60 | 459 | 0.19 | 0.16* | 349 |
| (Reconst. Sc.) | (0.07) | (0.08) | (0.07) | (14.33) | (570) | (0.65) | (0.09) | (726) |
| **Panel C**: HLATE, DD (Reconstructed Score) | | | | | | | | |
| Women | -0.01 | 0.30 | 0.18 | 32.57 | 1630 | 1.12 | 0.38 | 1220 |
| | (0.26) | (0.25) | (0.23) | (38.49) | (1604) | (2.05) | (0.27) | (1680) |
| Men | 0.13 | 0.50* | 0.48* | 76.71 | 1087 | 0.13 | 0.43 | 425 |
| | (0.21) | (0.28) | (0.26) | (58.26) | (1957) | (2.48) | (0.35) | (2318) |
| Difference | -0.14 | -0.20 | -0.29 | -44.15 | 542 | 0.99 | -0.05 | 795 |
| | (0.32) | (0.36) | (0.37) | (73.76) | (2408) | (3.38) | (0.42) | (2578) |
| Bandwidth | 10 | 10 | 10 | 10 | 10 | 10 | 10 | 10 |
| Observations | 621 | 621 | 621 | 621 | 590 | 590 | 590 | 336 |
| **Group means at Baseline** | **(Within a Given Bandwidth)** | | | | | | | |
| Women | 0.54 | 0.22 | 0.13 | 60.18 | 943 | 2.92 | 0.12 | 1736 |
| Men | 0.70 | 0.45 | 0.26 | 113.65 | 2074 | 4.73 | 0.33 | 2987 |
| **Panel D**: Heterogeneous ITT, DD (Reconstructed Score) | | | | | | | | |
| Women | -0.01 | 0.13 | 0.10 | 15.51 | 624 | 0.33 | 0.16 | 593 |
| | (0.09) | (0.09) | (0.08) | (14.54) | (623) | (0.77) | (0.10) | (871) |
| Men | 0.04 | 0.19* | 0.18** | 28.63 | 350 | 0.03 | 0.15 | 157 |
| | (0.07) | (0.10) | (0.09) | (21.19) | (658) | (0.83) | (0.12) | (781) |
| Difference | -0.05 | -0.06 | -0.08 | -13.12 | 274 | 0.30 | 0.01 | 436 |
| | (0.09) | (0.11) | (0.11) | (20.81) | (672) | (0.96) | (0.12) | (854) |

*Notes*: Standard errors adjusted for clustering at event level reported in parentheses. Each cell in Panel A and B represents an estimate from a separate regression, which includes the total score, the relative score (forcing variable), and an interaction of the relative score with the assignment variable as counterfactuals, as well as a constant. Panel C presents second stage results. For comparability, Panel B, C, and D estimates are conducted within the same bandwidths as in Panel A. ***, **, and * denote significance at the 1, 5, and 10 percent level.

**TABLE 10: TIME USE BY GENDER, SECOND FOLLOW-UP (2010 COHORT)**

| | **Hours/week working:** | | | | | |
|---|---|---|---|---|---|---|
| | Unpaid, inside house > 5 (1=Yes) | Unpaid, inside house > 10 (1=Yes) | Unpaid, inside house > 20 (1=Yes) | Unpaid, outside house > 0 (1=Yes) | Paid, inside house > 0 (1=Yes) | Paid, outside house > 0 (1=Yes) |
| | (1) | (2) | (3) | (4) | (5) | (6) |
| **Panel A**: HLATE, DD (Reconstructed Score) | | | | | | |
| Women | 0.20 | 0.32 | 0.30 | 0.43 | 0.26 | -0.09 |
| | (0.23) | (0.25) | (0.32) | (0.32) | (0.26) | (0.35) |
| Men | -0.24 | 0.23 | 0.26 | -0.09 | 0.13 | 0.22 |
| | (0.24) | (0.26) | (0.21) | (0.28) | (0.27) | (0.28) |
| Difference | 0.44 | 0.10 | 0.04 | 0.53 | 0.13 | -0.31 |
| | (0.37) | (0.34) | (0.35) | (0.43) | (0.33) | (0.41) |
| Bandwidth | 10 | 10 | 10 | 10 | 10 | 9 |
| Observations | 620 | 620 | 620 | 620 | 621 | 621 |
| **Group means at Baseline (Within a Given Bandwidth)** | | | | | | |
| Women | 0.95 | 0.82 | 0.54 | 0.55 | 0.41 | 0.43 |
| Men | 0.65 | 0.36 | 0.14 | 0.65 | 0.48 | 0.67 |
| **Panel B**: Heterogeneous ITT, DD (Reconstructed Score) | | | | | | |
| Women | 0.06 | 0.13 | 0.12 | 0.15 | 0.09 | -0.02 |
| | (0.08) | (0.09) | (0.12) | (0.11) | (0.11) | (0.13) |
| Men | -0.09 | 0.09 | 0.10 | -0.01 | 0.07 | 0.07 |
| | (0.09) | (0.09) | (0.08) | (0.10) | (0.10) | (0.10) |
| Difference | 0.15 | 0.04 | 0.03 | 0.15 | 0.03 | -0.09 |
| | (0.11) | (0.10) | (0.11) | (0.12) | (0.10) | (0.12) |

*Notes*: Standard errors adjusted for clustering at event level reported in parentheses. Panel A presents second stage results. In both panels, each cell represents a treatment effect retrieved from a regression, which includes the group variable, an interaction term of the group variable with the training or assignment indicator, the total score, the relative score (forcing variable), and an interaction of the relative score with the assignment variable as counterfactuals, as well as a constant. *Difference* is the coefficient of the interaction term with the group variable from the respective regression. ***, **, and * denote significance at the 1, 5, and 10 percent level.

### TABLE 11: SURVEY AND ADMINISTRATIVE DATA COMPARISION

| | EF Verification Data | EF Verification Data | Survey Data | EF Verification Data | Survey Data |
|---|---|---|---|---|---|
| **Months After Training** | 3 | 6 | 9-11 | 15 | 21-23 |
| **2010** | | | | | |
| IGA with Earnings > 3000 NRs. (1=Yes) | 79%[1] | - | 62% | - | 57% |
| **2011** | | | | | |
| Any IGA (1=Yes) | - | - | 83% | 72%[4] | - |
| IGA with Earnings > 3000 NRs. (1=Yes) | - | 80%[2] | 57% | - | - |
| **2012** | | | | | |
| Any IGA (1=Yes) | - | 91%[3] | 73% | - | - |
| IGA with Earnings > 3000 NRs. (1=Yes) | - | 81%[3] | 60% | - | - |
| IGA with Earnings > 4600 NRs. (1=Yes) | - | 67%[3] | 46% | - | - |

*Sources*:

[1] EF Annual Report 2010, p2.

[2] EF Annual Report 2011, p5.

[3] EF Annual Report 2012, p4.

[4] EF Tracer Study 2013, p23.

# APPENDIX

# A. Reconstruction of Assignment Score and Threshold

**FIGURE A1: APPLICATION FORM**

**Application Form**

Registration #: ______________

Stamp of the Training Institution

**Personal Details**

Name and Surname: ___________________________________________

Sex: ☐ Female ☐ Male ☐ Other

Marital Status: ☐ Single ☐ Married

Date of Birth (Day/Month/Year): ___________/____________/___________ Age: __________

Caste/Ethnicity: ☐ Dalit ☐ Janajati ☐ Others (please specify)______________

Special circumstances: ☐ HIV-infected ☐ Disabled ☐ Widow ☐ Ex-combatants ☐ Internally Displaced People

| Permanent Address | |
|---|---|
| District: | |
| Municipality/ VDC: | |
| Tole: | Ward No: |
| Phone Number (Land line): | |
| Phone Number (Mobile): | |
| Father's Name: | |
| Mother's Name: | |
| Citizenship No.: | |
| Issued District: | |

| Current Address | |
|---|---|
| District: | |
| Municipality/ VDC: | |
| Tole: | Ward No: |
| Phone Number (Land line): | |
| Phone Number (Mobile): | |

**In case of getting information of you:**

| | |
|---|---|
| Reference person: | |
| Mobile Number: | |

**Education Details**

**Highest completed level of education:**

☐ Illiterate ☐ Below class 5 ☐ Class 5-8 ☐ Class 9-10 ☐ SLC Pass ☐ +2 ☐ Bachelor's degree and above

*If you have an academic certificate, please attach a copy.*

**Employment and Income Information**

**What is your employment status and monthly earnings (own earnings only):**

☐ Self-employed Monthly earnings:____________ ☐ Wage earner Monthly earnings:____________

☐ Agriculture Monthly earnings:____________ ☐ Unemployed Monthly earnings:____________

☐ Other ___________ Monthly earnings:____________

**Estimated total monthly income Rs. __________**

**What is your family's average annual earnings in the following areas (excluding your own)?**

☐ Labour Wages __________ ☐ Salary __________ ☐ Business __________

☐ Animal Husbandry __________ ☐ Foreign income __________ ☐ Others __________

**Number of family members __________**

Estimated total monthly income per member Rs. __________

*(Please divide the sum of total annual income by total no of family members and 12 months)*

**FIGURE A1: APPLICATION FORM (CONTINUED)**

**How many months can you feed your family through agricultural income from your own land?**

☐ Less than 3 months ☐ Less than 6 months ☐ More than 6 months ☐ Do not have land

**Training Information**

Name of training:_______________________________________________

District: ____________ V.D.C: ____________ Ward No.: ____________ Start date:____________

**Reason for interest in training:**

☐ To start own work ☐ To work for wages
☐ To upgrade skills ☐ To go for foreign employment

**How did you come to know about this training?**

☐ Newspaper ☐ Relatives/Friends ☐ Muslim Women Society
☐ Poster ☐ Training Centre ☐ Dalit Women Association
☐ Pamphlets ☐ Local Development Agencies ☐ Indigenous People Association
☐ FM radio ☐ Single Women Group ☐ Other ________________

**Have you had any previous training?** ☐ Yes ☐ No

**If yes, please provide the following information:**

Name of Training:_______________________________________________

Hours of Training:_______________________________________________

Date completed:_______________________________________________

I state that the above-mentioned details are true.

____________________ ____________________
Signature of Applicant Date (Day/Month/Year)

## FIGURE A2: SAMPLE RANKING FORM

| # | Name and Surname | Immediate contact telephone | Entry Requirement (Y/N) | | | Selection Criteria (Individual Scores): 1-4. Short-listing (70%) | | | | Selection Criteria (Individual Scores): 5. Interview (30%) | Final Score | Rank |
|---|---|---|---|---|---|---|---|---|---|---|---|---|
| | | | Age 16-35 for men, 16-40 for women *(Write age)* | Education < SLC (completion of 10th grade) | Less than 6 months food sufficiency (rural areas) or income < NRs. 3,000 | 1. Trade-specific education (15) | 2. Economic status (20) | 3. Social caste (25) | 4. Development status of district of origin (10) | | **TOTAL** SCORE (100) *List candidates from highest to lowest* | |
| 1 | Jane Doe 1 | 12345678 | 28 | Y | Y | 15 | 20 | 20 | 10 | 27 | 92 | 1 |
| 2 | Jane Doe 2 | 12345678 | 29 | Y | Y | 15 | 20 | 20 | 10 | 25 | 90 | 2 |
| 3 | John Doe 1 | 12345678 | 26 | Y | Y | 15 | 20 | 20 | 10 | 24 | 89 | 3 |
| 4 | Jane Doe 3 | 12345678 | 20 | Y | Y | 15 | 20 | 20 | 0 | 30 | 85 | 4 |
| 5 | Jane Doe 4 | 12345678 | 21 | Y | Y | 15 | 20 | 20 | 5 | 20 | 80 | 5 |
| 6 | John Doe 2 | 12345678 | 24 | Y | Y | 15 | 15 | 20 | 5 | 25 | 80 | 6 |
| 7 | Jane Doe 5 | 12345678 | 19 | Y | Y | 15 | 20 | 15 | 0 | 29 | 79 | 7 |
| 8 | Jane Doe 6 | 12345678 | 33 | Y | Y | 15 | 15 | 25 | 10 | 13 | 78 | 8 |
| 9 | John Doe 3 | 12345678 | 17 | Y | Y | 10 | 15 | 20 | 5 | 28 | 78 | 9 |
| 10 | Jane Doe 7 | 12345678 | 21 | Y | Y | 15 | 20 | 15 | 5 | 22 | 77 | 10 |
| 11 | Jane Doe 8 | 12345678 | 27 | Y | Y | 15 | 15 | 10 | 10 | 26 | 76 | 11 |
| 12 | John Doe 4 | 12345678 | 23 | Y | Y | 15 | 20 | 10 | 10 | 20 | 75 | 12 |
| 13 | Jane Doe 9 | 12345678 | 18 | Y | Y | 15 | 15 | 20 | 0 | 25 | 75 | 13 |
| 14 | Jane Doe 10 | 12345678 | 35 | Y | Y | 15 | 15 | 20 | 0 | 23 | 73 | 14 |
| 15 | John Doe 5 | 12345678 | 19 | Y | Y | 15 | 15 | 20 | 5 | 18 | 73 | 15 |
| 16 | Jane Doe 11 | 12345678 | 22 | Y | Y | 15 | 0 | 20 | 10 | 27 | 72 | 16 |
| 17 | Jane Doe 12 | 12345678 | 30 | Y | Y | 5 | 20 | 25 | 0 | 16 | 66 | 17 |
| 18 | John Doe 6 | 12345678 | 25 | Y | Y | 15 | 10 | 20 | 5 | 15 | 65 | 18 |
| 19 | Jane Doe 13 | 12345678 | 24 | Y | Y | 15 | 15 | 10 | 5 | 10 | 55 | 19 |
| 20 | John Doe 7 | 12345678 | 32 | Y | Y | 15 | 15 | 10 | 5 | 6 | 51 | 20 |

*Notes*: Red line indicates cut-off between accepted and rejected candidates. Candidates are sampled for the survey if their score is within a range of the cut-off score plus/minus 20 percent. The shaded area represents candidates who would have been sampled for the baseline survey based on this example.

**FIGURE A3: EFFECT SIZES BY DIFFERENT ABSOLUTE THRESHOLD VALUES**

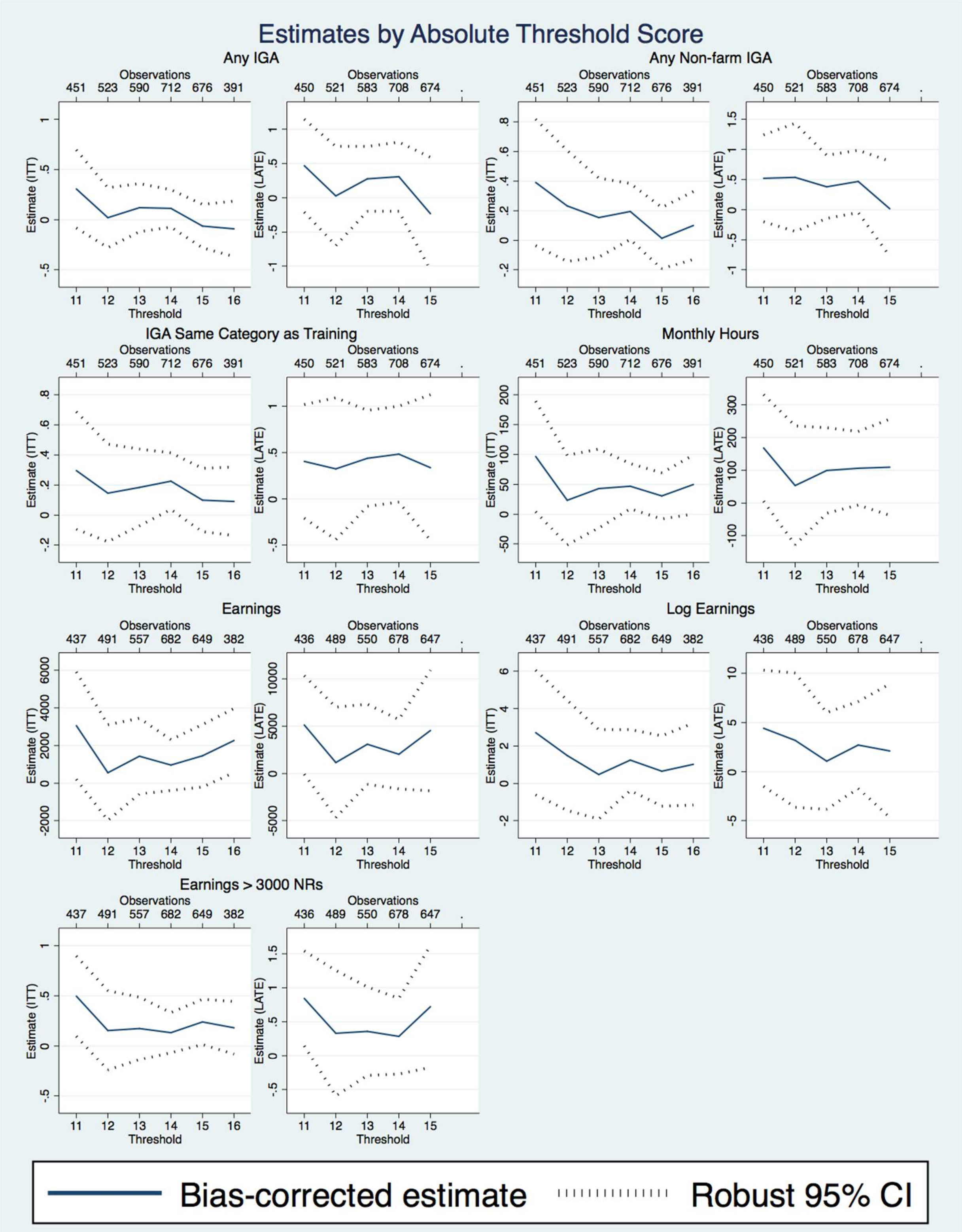

Graph reports estimates by groups of absolute threshold at course level. Only those estimates are reported which could be based on a sample size of at least 300 observations for ITT and 400 observations for LATE. Conditional earnings, non-farm wage employment, and non-farm self-employment are not reported as most bins feature less than 300 observations. We apply the user-written program rdrobust provided by Calonico, Cattaneo, and Titiunik (2014) to obtain bias-corrected estimates and robust confidence intervals.

**FIGURE A4: CROSS-VALIDATION FUNCTION FOR OUTCOMES IN LEVELS**

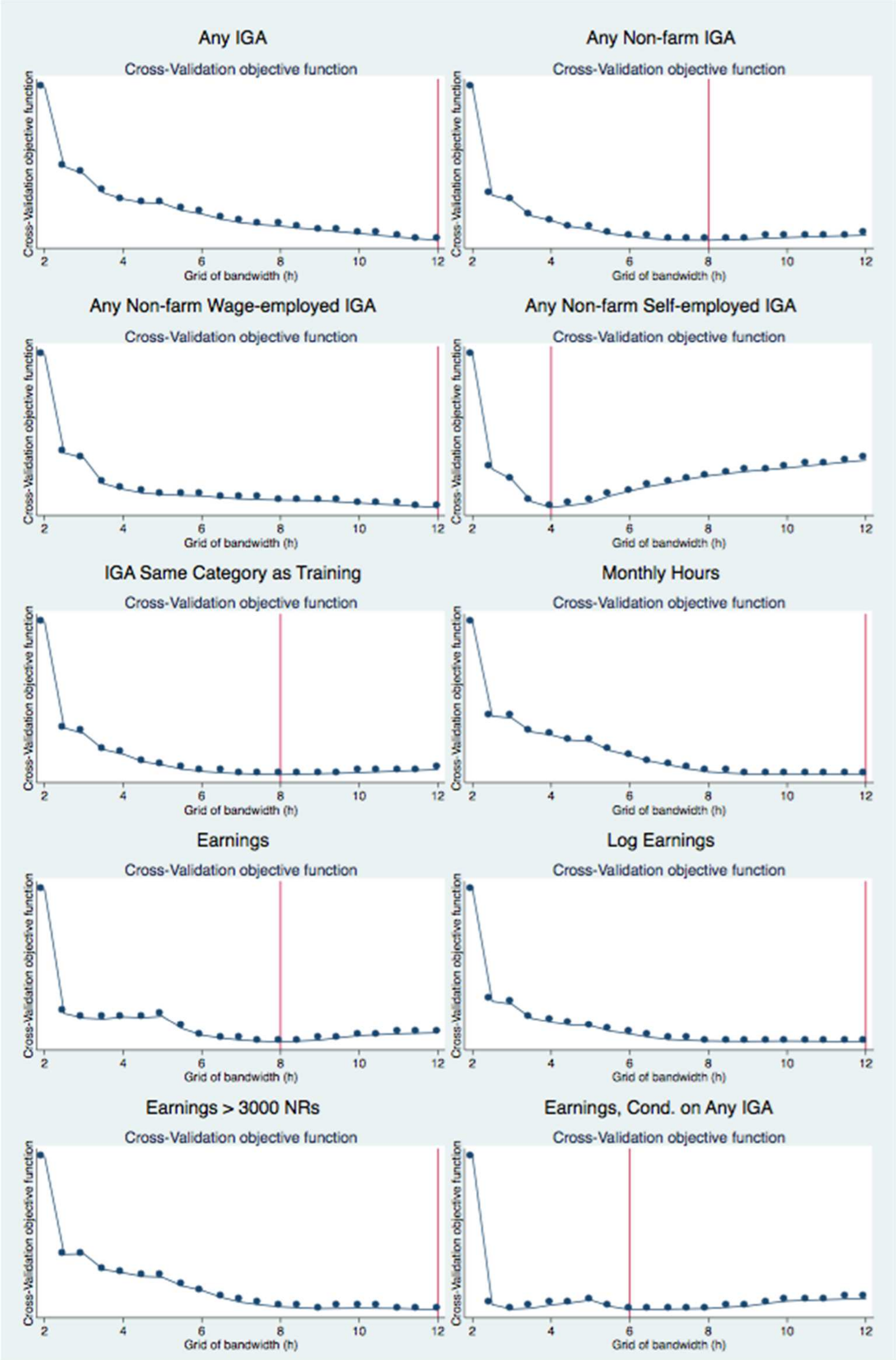

**FIGURE A5: CROSS-VALIDATION FUNCTION FOR DIFFERENCED OUTCOMES**

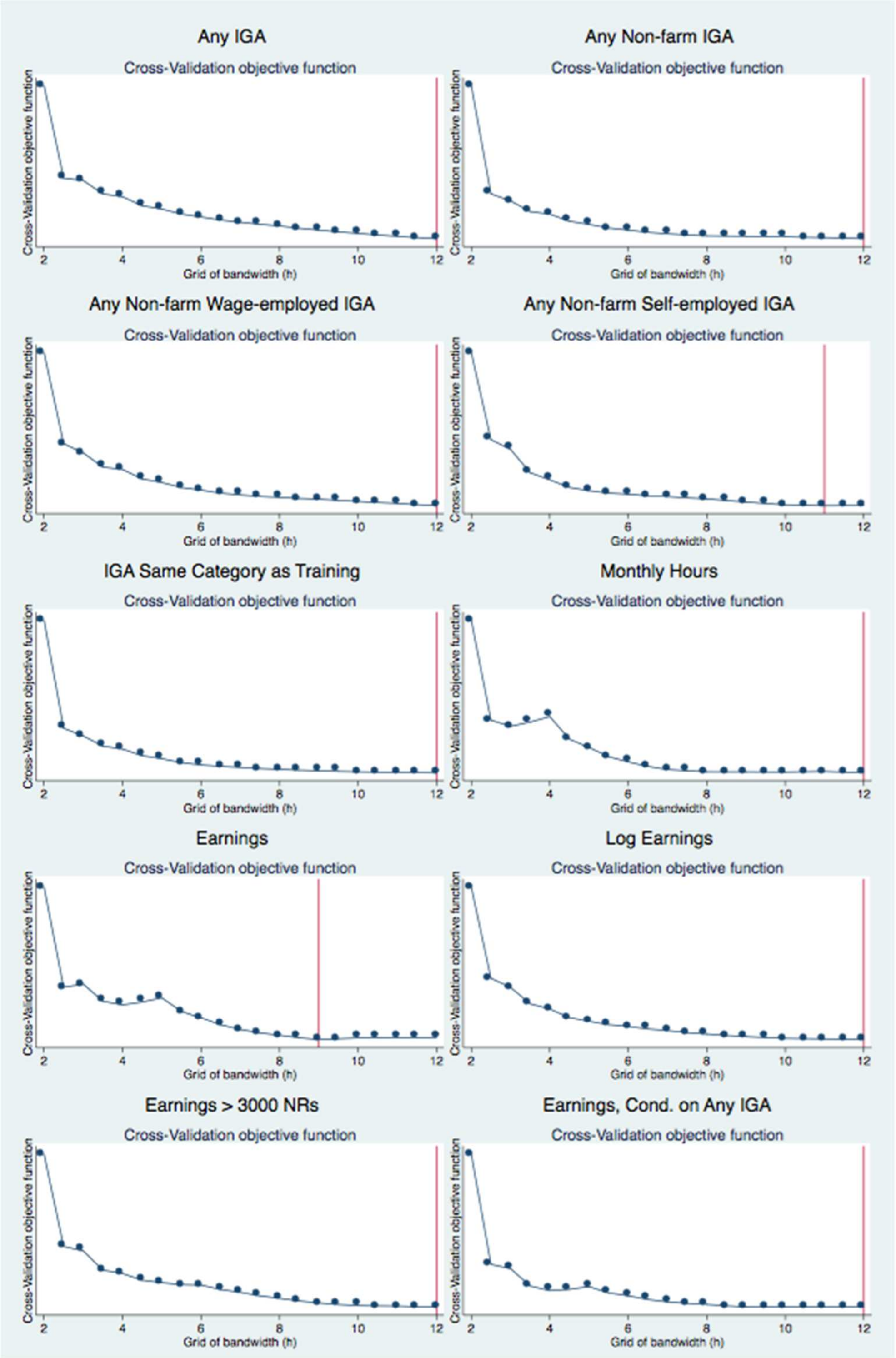

TABLE A1: ORIGINAL AND RECONSTRUCTED SCORE COMPONENTS

| Sub-Score | Total weight | Basis for evaluation | Indicators | Original Score | Reconstructed Score | |
|---|---|---|---|---|---|---|
| | | | | Assigned Sub-Score | Assigned Sub-Score | Source |
| 1 | 15% | Trade-specific education requirement | Compulsory prerequisite: All candidates must meet the minimum requirement for the training they applied to* | 15<br><br>(0, 5 or 10 in exceptional cases) | Integers of 0 to 3 | Predicted |
| 2 | 20% | Economic poverty | Less than 3 months of food sufficiency | 20 | 4 | Application form data |
| | | | Less than 6 months of food sufficiency or less than 3000 per capita family income from non-farm based income | 15 | 3 | |
| | | | More than 6 months of food sufficiency and per capita family income from non-farm based income equal to or more than 3000 | 0 | 0 | |
| 3 | 25% | Social caste | Women: Dalit women or women from the following special groups: widows; internally displaced; ex-combatants; physically disabled; HIV-infected infected | 25 | 5 | Application form data |
| | | | Women: Economically poor women not referred to above | 20 | 4 | |
| | | | Men: Dalit, Janjati, Madhesi men or men from the following special groups: internally displaced; ex-combatants; physically disabled; HIV-infected infected | 15 | 3 | |
| | | | Men: Economically poor men not referred to above | 10 | 2 | |
| | | | Neither of the above | 0 | 0 | |
| 4 | 10% | Development status of district of origin | Least developed districts | 10 | 2 | Application form data |
| | | | Moderately developed districts | 5 | 1 | |
| | | | Developed districts | 0 | 0 | |
| **Preliminary score for short-listing (Sub – total)** | | | | **70** | **14** | |
| 5 | 30% | Interview | Commitment, Motivation, Attitude, Aptitude, Clear Vision for Employment and Enterprising | 0-30 | 0-6 | Predicted |
| **Total score after interview** | | | | **100** | **20** | |

*Notes*: *If candidates did not fulfill the course specific education prerequisites they were not eligible for short-listing and immediately rejected. In exceptional cases (approx. 9 percent of the sample) this criterion was not adhered to and instead applicants received 0, 5, or 10 points. When reconstructing this component, we therefore allow for integer values between 0 and 3.

**TABLE A2: DEVELEOPMENT STATUS OF NEPAL DISTRICTS**

| Developed District | Moderately Developed District | Least Developed District |
|---|---|---|
| Kathmandu | Makawanpur | Ramechhap |
| Chitwan | Gulmi | Parsa |
| Jhapa | Surkhet | Rasuwa |
| Bhaktapur | Solukhumbu | Kapilbastu |
| Lalitpur | Banke | Bara |
| Kaski | Bhojpur | Dadeldhura |
| Dhankuta | Gorkha | Darchula |
| Palpa | Taplejung | Siraha |
| Syangja | Bardiya | Jajarkot |
| Manang | Kanchanpur | Rukum |
| Morang | Nuwakot | Sarlahi |
| Illam | Nawalparasi | Baitadi |
| Rupandehi | Khotang | Dailekh |
| Sunsari | Okhaldhunga | Rolpa |
| Kabhreplanchok | Kailali | Mahotari |
| Tanahu | Dolakha | Doti |
| Terhathum | Arghakhanchi | Dolpa |
| Sankhuwasabha | Udayapur | Rautahat |
| Mustang | Dhading | Jumla |
| Parbat | Salyan | Kalikot |
| Dang | Dhanusa | Bajura |
| Lamjung | Saptari | Achham |
| Panchthar | Sindhipalchok | Bajhang |
| Baglung | Sundhuli | Humla |
| Myagdi | Pyuthan | Mugu |

*Source*: Districts of Nepal, Indicators of Development. Updated 2003. CBS/Nepal, ICIMOD. December, 2003

**TABLE A3: CORRELATION OF 5$^{TH}$ SUB-SCORE WITH OTHER SCORE COMPONENTS**

| | Sub-Score 5<br>(1) | Sub-Score 5<br>(2) | Sub-Score 5<br>(3) |
|---|---|---|---|
| Aggregated Sub-Scores 1 to 4 | -0.220***<br>(0.021) | | |
| Sub-Score 1 (Trade-specif. Edu.) | | 0.117*<br>(0.062) | 2.752***<br>(0.431) |
| Sub-Score 2 (Econ. Poverty) | | -0.182***<br>(0.029) | 1.046*<br>(0.551) |
| Sub-Score 3 (Social Caste) | | -0.348***<br>(0.031) | 1.013***<br>(0.350) |
| Sub-Score 4 (Development Status of Dist. Of Origin) | | -0.473***<br>(0.095) | 3.560***<br>(1.354) |
| SS 1 x SS 2 | | | -0.120***<br>(0.039) |
| SS 1 x SS 3 | | | -0.136***<br>(0.027) |
| SS 1 x SS 4 | | | -0.379***<br>(0.101) |
| SS 2 x SS 3 | | | -0.049<br>(0.030) |
| SS 2 x SS 4 | | | -0.194**<br>(0.096) |
| SS 3 x SS 4 | | | -0.278***<br>(0.087) |
| SS 1 x SS 2 x SS 3 | | | 0.006***<br>(0.002) |
| SS 1 x SS 2 x SS 4 | | | 0.019***<br>(0.007) |
| SS 1 x SS 3 x SS 4 | | | 0.026***<br>(0.006) |
| SS 2 x SS 3 x SS 4 | | | 0.013**<br>(0.006) |
| SS 1 x SS 2 x SS 3 x SS 4 | | | -0.001***<br>(0.000) |
| N | 4090 | 4090 | 4090 |
| Adj. R2 | 0.38 | 0.39 | 0.40 |

*Notes*: All models include event dummies. ***, **, and * denote significance at the 1, 5, and 10 percent level, respectively.

**TABLE A4: ATTRITION**

| | Pooled Cohorts, 1st Follow-Up | | | | 2010 Cohort, 2nd Follow-Up | | | |
|---|---|---|---|---|---|---|---|---|
| | (1) | (2) | (3) | (4) | (5) | (6) | (7) | (8) |
| "Above Threshold" | 0.008 | 0.017 | 0.007 | 0.016 | -0.011 | -0.047 | -0.004 | -0.052 |
| | (0.010) | (0.019) | (0.010) | (0.019) | (0.027) | (0.045) | (0.029) | (0.047) |
| Female X "Above Threshold" | | -0.020 | | -0.014 | | 0.059 | | 0.080 |
| | | (0.021) | | (0.021) | | (0.066) | | (0.066) |
| Female | | 0.077*** | 0.066*** | 0.074*** | | -0.026 | 0.026 | -0.019 |
| | | (0.021) | (0.014) | (0.020) | | (0.060) | (0.035) | (0.056) |
| Age | | | 0.004*** | 0.004*** | | | 0.007** | 0.007** |
| | | | (0.001) | (0.001) | | | (0.003) | (0.003) |
| Parent | | | -0.003 | -0.003 | | | -0.006 | -0.009 |
| | | | (0.019) | (0.019) | | | (0.045) | (0.045) |
| Married | | | 0.011 | 0.011 | | | 0.006 | 0.010 |
| | | | (0.019) | (0.019) | | | (0.043) | (0.043) |
| Dalit | | | -0.052** | -0.052** | | | -0.052 | -0.053 |
| | | | (0.023) | (0.023) | | | (0.062) | (0.062) |
| Janjati | | | -0.014 | -0.014 | | | 0.109*** | 0.110*** |
| | | | (0.012) | (0.012) | | | (0.032) | (0.032) |
| Any IGA at baseline | | | 0.017 | 0.017 | | | 0.024 | 0.023 |
| | | | (0.011) | (0.011) | | | (0.027) | (0.027) |
| N | 4585 | 4585 | 4585 | 4585 | 1547 | 1547 | 1547 | 1547 |
| Training provider dummies | Yes | Yes | Yes | Yes | Yes | Yes | Yes | Yes |
| District dummies | Yes | Yes | Yes | Yes | Yes | Yes | Yes | Yes |

*Notes*: All models include district and training provider dummies. In Columns (3), (4), (7), and (8) indicators are included which flag missing values for each of the control variables, respectively. All models are estimated based on the reconstructed score. The difference in sample size between the initial baseline sample and the sample we use in this analysis arises due to missing values in the variables that were necessary to reconstruct the score variable, which determines assignment. All standard errors are clustered at event level. ***, **, and * denote significance at the 1, 5, and 10 percent level, respectively.

**TABLE A5: BANDWIDTH SENSITIVITY**

| | Any IGA (1=Yes) | Any non-farm IGA (1=Yes) | Non-farm wage IGA (1=Yes) | Non-farm self IGA (1=Yes) | Trade-specific IGA (1=Yes) | Hours worked in past month | Earnings | Logged earnings | Earnings > 3000 NRs. (1=Yes) | Earnings, conditional on any IGA |
|---|---|---|---|---|---|---|---|---|---|---|
| | (1) | (2) | (3) | (4) | (5) | (6) | (7) | (8) | (9) | (10) |
| **Panel A**: LATE, DD (Reconstructed Score) | | | | | | | | | | |
| 2 Index Scores | 0.19 | 0.50** | 0.35 | 0.14 | 0.43** | 121.8*** | 3387** | 2.33 | 0.43** | 3058 |
| | (0.21) | (0.20) | (0.31) | (0.27) | (0.20) | (42.98) | (1341) | (1.63) | (0.21) | (2254) |
| Observations | 2133 | 2133 | 1364 | 1364 | 2133 | 2133 | 2049 | 2049 | 2049 | 1012 |
| F-statistic | 27.50 | 27.50 | 12.22 | 12.22 | 27.50 | 27.50 | 27.83 | 27.83 | 27.83 | 12.85 |
| 3 Index Scores | 0.13 | 0.38** | 0.28 | 0.18 | 0.45*** | 77.66** | 2319** | 1.52 | 0.34** | 1722 |
| | (0.16) | (0.15) | (0.19) | (0.17) | (0.16) | (31.35) | (1083) | (1.27) | (0.16) | (1645) |
| Observations | 2874 | 2874 | 1862 | 1862 | 2874 | 2874 | 2759 | 2759 | 2759 | 1411 |
| F-statistic | 36.18 | 36.18 | 20.31 | 20.31 | 36.18 | 36.18 | 37.95 | 37.95 | 37.95 | 17.43 |
| 4 Index Scores | 0.09 | 0.31** | 0.15 | 0.25* | 0.45*** | 58.86** | 1755** | 1.26 | 0.24** | 1258 |
| | (0.12) | (0.12) | (0.13) | (0.14) | (0.13) | (24.97) | (872) | (1.06) | (0.12) | (1097) |
| Observations | 3340 | 3340 | 2145 | 2145 | 3340 | 3340 | 3202 | 3202 | 3202 | 1622 |
| F-statistic | 62.42 | 62.42 | 33.90 | 33.90 | 62.42 | 62.42 | 65.38 | 65.38 | 65.38 | 37.29 |
| 5 Index Scores | 0.07 | 0.29*** | 0.10 | 0.25** | 0.41*** | 52.67** | 1666** | 1.29 | 0.24** | 1393 |
| | (0.11) | (0.10) | (0.12) | (0.12) | (0.12) | (20.54) | (744) | (0.92) | (0.10) | (994) |
| Observations | 3685 | 3685 | 2391 | 2391 | 3685 | 3685 | 3530 | 3530 | 3530 | 1778 |
| F-statistic | 77.89 | 77.89 | 42.88 | 42.88 | 77.89 | 77.89 | 79.88 | 79.88 | 79.88 | 42.02 |
| 10 Index Scores | 0.11 | 0.29*** | 0.09 | 0.28** | 0.41*** | 45.13** | 1685** | 1.53* | 0.28*** | 1662 |
| | (0.10) | (0.10) | (0.11) | (0.12) | (0.12) | (20.99) | (705) | (0.89) | (0.11) | (1017) |
| Observations | 4000 | 4000 | 2604 | 2604 | 4000 | 4000 | 3834 | 3834 | 3834 | 1929 |
| F-statistic | 83.98 | 83.98 | 51.84 | 51.84 | 83.98 | 83.98 | 82.98 | 82.98 | 82.98 | 38.61 |
| **Panel B**: ITT, DD (Reconstructed Score) | | | | | | | | | | |
| 2 Index Scores | 0.06 | 0.15** | 0.08 | 0.03 | 0.13** | 36.23*** | 1033*** | 0.71 | 0.13** | 885 |
| | (0.06) | (0.06) | (0.07) | (0.06) | (0.06) | (12.15) | (392) | (0.51) | (0.06) | (602) |
| 3 Index Scores | 0.03 | 0.11** | 0.07 | 0.05 | 0.13*** | 21.29** | 662** | 0.42 | 0.10** | 455 |
| | (0.04) | (0.04) | (0.05) | (0.04) | (0.05) | (8.66) | (309) | (0.38) | (0.04) | (437) |
| 4 Index Scores | 0.03 | 0.10** | 0.05 | 0.07* | 0.15*** | 18.97** | 587** | 0.42 | 0.08** | 435 |
| | (0.04) | (0.04) | (0.04) | (0.04) | (0.04) | (8.26) | (295) | (0.37) | (0.04) | (386) |
| 5 Index Scores | 0.02 | 0.10*** | 0.03 | 0.08** | 0.14*** | 17.88** | 579** | 0.45 | 0.09** | 473 |
| | (0.04) | (0.04) | (0.04) | (0.04) | (0.04) | (7.20) | (264) | (0.33) | (0.04) | (344) |
| 10 Index Scores | 0.03 | 0.09*** | 0.03 | 0.08** | 0.13*** | 14.11** | 532** | 0.49* | 0.09*** | 494 |
| | (0.03) | (0.03) | (0.03) | (0.03) | (0.04) | (6.60) | (223) | (0.29) | (0.03) | (305) |

*Notes*: Standard errors adjusted for clustering at event level reported in parentheses. Panel A presents second stage results. In both panels, each cell represents an estimate from a separate regression, which includes the total score, the relative score (forcing variable), and an interaction of the relative score with the assignment variable as counterfactuals, as well as a constant. *Non-farm wage IGA* and *Non-farm self IGA* are only available for 2011 and 2012 cohorts. ***, **, and * denote significance at the 1, 5, and 10 percent level.

## B. Robustness Tests Based on Propensity Score Techniques

We estimate:

$$Y_{it} = \alpha + \delta\,(A_{it}\,x\,Treat_{it}) + u_i + v_t + \varepsilon_{it} \qquad \text{(B1)}$$

$Y_{it}$ is the employment outcome of interest for individual $i$ from training event $j$ at time $t$; $Treat_i$ is an indicator, equal to 1 for the treatment group and 0 for control; $A_{it}$ is an indicator for the period when treatment occurs; $u_i$ captures program effects; $v_t$ captures the time effects, $\varepsilon_{it}$, is an idiosyncratic error term, clustered by training event. After estimating a propensity score[35], we derive the estimated treatment effect using two methods: "inverse propensity score weighting" (IPSW) and nearest neighbor matching (NN). In the IPSW method individuals are weighted according to the inverse of their estimated propensity to participate in the program. The weighted observations are then used in a DID regression. We present this method estimates under the IPSW specification in the Appendix B tables.[36,37] The NN matching algorithm, in which each individual in the treatment group is compared to a fixed number of control observations (in our estimation we use four observations) with the closest propensity score. We present "NN specification" in Appendix B. Following Smith and Todd (2005), we estimate the difference-in-difference matching estimator for the training program effect $\delta$ as follows:

$$\widehat{\delta_M} = \frac{1}{N_T} \sum_{i \epsilon T} \left[ \left( y_{it_1} - y_{it_0} \right) - \sum_{j \in C} W_{ij} \left( y_{jt_1} - y_{jt_0} \right) \right] \qquad \text{(B2)}$$

$N_T$ is the number of treatment observations, the subscript $t_1$ denotes follow-up observations and $t_0$ denotes baseline observations; $W_i$ is a matrix of weights. Weights for nearest-neighbor matching are computed by:

[35] We employ various specifications, including the individual training score, the individual training score and demographic variables, the five subinterview scores and finally the demographic variables plus provider/district/cohort/city fixed effects. The results are stable across various specifications though we report the last specification based on demographic variables plus provider/district/cohort/city fixed effects because that specification yields the best overlap of treatment and control unit distributions in the common support area.

[36] We implement IPSW following Hirano et al. (2003).

[37] This particular weighting method, as opposed to matching approaches, has the nice property of including all the data (unless weights are set to 0) and does not depend on random sampling, thus providing for replicability. We use a weighted least squares regression model, with weights of $1/\hat{\pi}$ for the treatment group and $1/(1-\hat{\pi})$ for the control group, where $\hat{\pi}$ is the estimated propensity score from (2). Standard errors are clustered by training event.

$$W_{ij}\left(y_{jt_1} - y_{jt_0}\right) = \frac{1}{x}\sum_{\substack{j=1 \\ j\in A_x}}^{x}\left(y_{jt_1} - y_{jt_0}\right) \quad (5)$$

$A_x$ is a set of $x$ observations with the lowest values of $\left|\hat{\pi}_i - \hat{\pi}_j\right|$. As in the two previous models outlined in this section, the dependent variable is the first difference of a given outcome between the baseline observation and the follow-up observation. We measure outcomes approximately one year after the start of training.[38, 39]

Appendix Table B11 shows the ATT results on employment and earnings for the pooled 2010, 2011 and 2012 cohorts based on the combined difference-in-difference and propensity score matching techniques. Unlike the RDD results, in this specification, we detect strong evidence of consistent impact on the employment rate across all specifications.[40] All three models indicate a positive and significant effect, despite the high employment rate (i.e., 61 percent) at baseline. Restricting the employment to non-farm activities, we also find a significant increase: the rate of participation in non-farm income-generating activities increases by 21 percentage points (from a base of 29.6 percent). Translating the results in percentage change terms, we find that the program increased non-farm employments by 71 percent. These impacts are not only statistically significant but also economically meaningful. We detect strong program impacts, though smaller in impacts than revealed by the RDD approach, on monthly earnings. We observe a statistically significant (at the 1 percent level) increase in monthly earnings for the treatment group by 976 to 1099 NRs (≈ 14 USD), from a baseline average of 1272 NRs (≈ 17 USD).[41] In percentage terms, this earnings increase translates to a 81 percent for the pooled sample. The impact on logged earnings is a little over 100 percent. The EF training program increased the "gainful employment" rate (i.e., the rate of new employment with earnings over 3000) increases by 16 to 17 percentage points, a result statistically significant across all three models. We also examine the trade-specific income generating activity (IGA) rate – the percent of individuals who find employment in the same trade as the training that they applied for – and we find impacts of 24 percentage points. The trade-specific IGA impacts are larger than the non-farm employment impacts, suggesting that members of the control group, even when able to find employment, were less able than the treatment group to find employment in the trade in which they sought training. Based on the propensity score approach, we find that the EF program leads to persistent improvements in the underemployment rate (i.e., cases in which people are working fewer hours than they wish). Table B11 shows that EF-sponsored training courses increased hours worked in IGAs for the pooled cohorts by 30-31 hours per month (i.e., 44 percent). All three model specifications exhibit a statistically significant and positive impact.

[38] Because the EF-sponsored training courses vary in length from 1 to 3 months, the follow-up survey examines outcomes 9 to 11 months after the end of the training.

[39] First, we address concerns about pre-existing differences and time-varying trends that could account for observed training effects when comparing trainees and non-trainees.[39] Table B1 presents baseline participant characteristics (i.e., balancing tests) for a set of 38 demographic indicators. These tests are based on "ITT" comparisons of the treatment group (i.e., individuals whose scores qualify them for admission to an EF training event) and the control group. The baseline balance tests for the pooled sample (2010-2012) indicate that significant differences exist between treatment and control groups for baseline observable characteristics and pre-treatment outcome variables.[39] Relative to rejected candidates, treated individuals are more likely to be Janajati and are less likely to have finished SLC (10th grade), characteristics which reflect the eligibility criteria and the EF"s differential pricing scheme for vulnerable groups. Furthermore, the likelihood of treated individuals being engaged in non-farm and trade specific employment before take up of training was higher, as well as their working hours and ability to earn more than 3,000 NRs a month. These differences are consistent with training providers' incentives to select candidates they think will perform best. Finally, individuals in the treatment group are also less likely to have control over savings and money of their own at baseline. To address these differences (and potential differences in unobservable characteristics) we applied a difference-in-difference approach in our analysis. However, growth in outcome variables and the may not follow a common trend, particularly when starting off at very different initial levels. Although it does not resolve the parallel trend assumption, we additionally applied propensity score weighting and matching techniques to achieve a higher degree of baseline comparability across groups.

[40] We measure employment by whether the respondent reported any income-generating activities in the past month or not.

[41] This average is based on the entire study cohort, including those with zero earnings at baseline. The average earnings among those with non-zero earnings were 2,928 NRs, translating to a percentage increase in earnings of 30 percent.

Table B12 shows results for program impact heterogeneity. The impacts of the skills training program differed markedly by type of trade for the pool 2010-2012 samples. Consistent and exactly aligned with the results based on the RDD approach, training in electronics, beautician, and tailoring consistently show strong ATT impacts on employment—graduates of these training programs are more likely to have employment in general and are also more likely to be working in the trade in which they were trained. Beautician training shows large impacts on both employment and earnings. We detect no significant impacts on employment or earnings outcomes for the remaining four trades. Results for food and hospitality (e.g., cooking and wait service) show no significant ATT impacts; results for construction show no significant impacts except for a marginal impact on trade-specific employment and on earning more than 3000 NRs per month. For the remaining three trades (i.e., poultry technician, handicrafts and farming), we detect some ATT impacts but they are not consistent across models. Overall, the results in Tables B12 reveal substantial heterogeneity in employment outcomes across the various types of training. The positive and significant impacts are driven almost entirely by three categories of trades: electronics, beautician training, and tailoring trades show positive and significant impacts on employment and earnings across both cohorts. We find no impacts for the food/hospitality and farming training. Construction-related trainings showed positive and significant impacts, but the effects are not consistent across outcomes.

Finally, we show program impacts for men and women (shown in Table B13). To that end, Tables B13-B14 disaggregate the results to compare outcomes for men versus women, and for younger women (the "AGEI" population) versus older women. Corroborating the RDD findings,

Tables B13-B14 show the employment impacts are larger, almost double the magnitude for women than they are for men. The results for other economic outcomes, such as hours worked, earnings, and type of employment, are similar for both sexes.

**FIGURE B1: DISTRIBUTION OF ESTIMATED PROPENSITY SCORES FOR 2010-2012 POOLED COHORTS (ITT)**

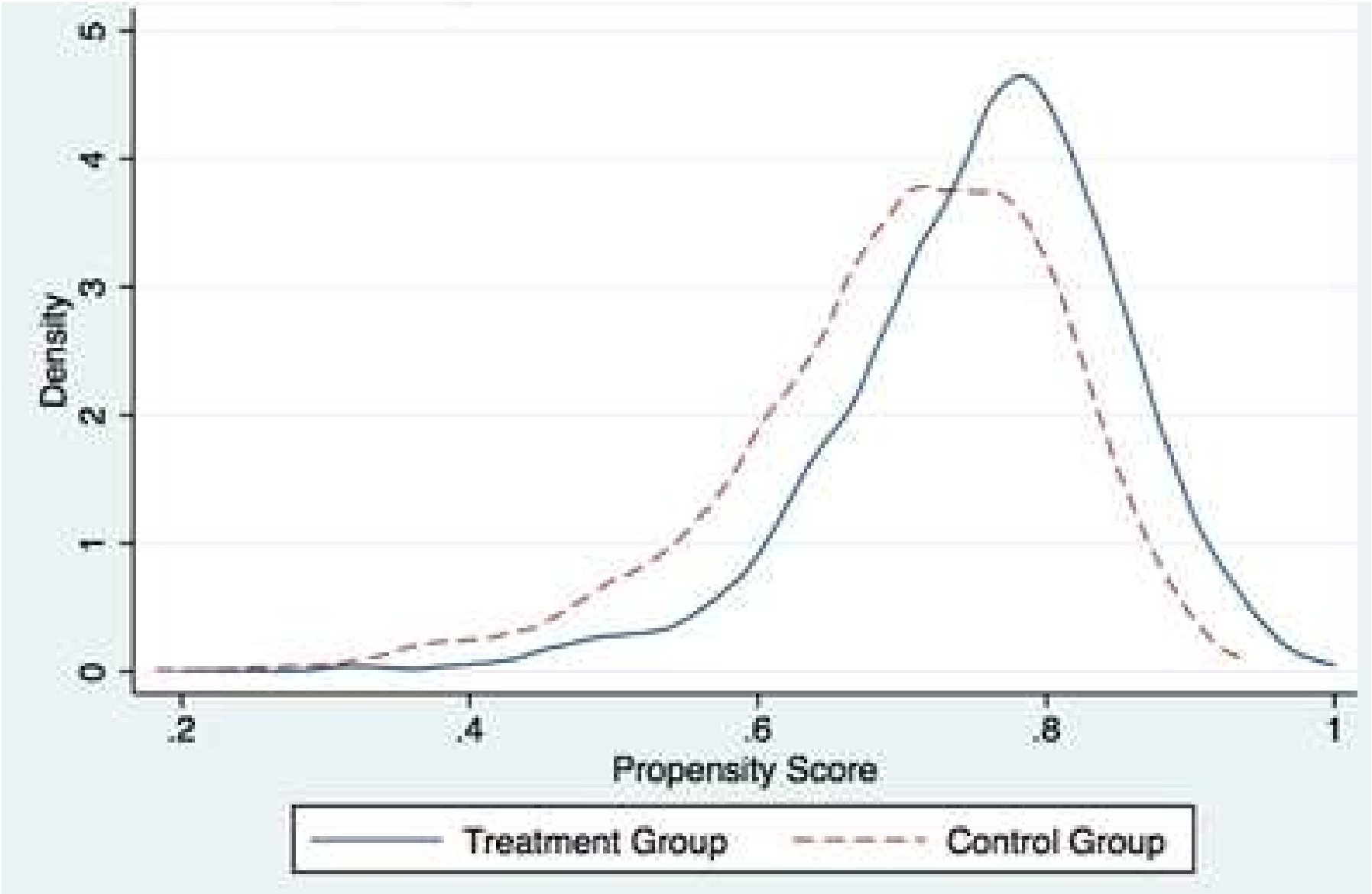

*Notes*: Propensity Score Distributions (Baseline ITT)

**TABLE B1: BASELINE BALANCING TESTS 2010-2012 POOLED (ITT), FULL SAMPLE**

| | Control | Treatment | Difference | p-value | N |
|---|---|---|---|---|---|
| *Demographics* | | | | | |
| Female (%) | 0.640 | 0.630 | -0.010 | 0.610 | 4101 |
| AGEI (women aged 16-24) (%) | 0.319 | 0.336 | 0.017 | 0.350 | 4101 |
| Dalit (%) | 0.090 | 0.077 | -0.012 | 0.365 | 4037 |
| Janajati (%) | 0.421 | 0.468 | 0.048** | 0.024 | 4037 |
| Muslim (%) | 0.017 | 0.025 | 0.008 | 0.269 | 4037 |
| Age | 24.537 | 24.242 | -0.294 | 0.249 | 4101 |
| Currently Married (%) | 0.580 | 0.594 | 0.014 | 0.463 | 4101 |
| Any Children (%) | 0.505 | 0.526 | 0.021 | 0.248 | 4101 |
| Completed SLC (10th grade) (%) | 0.163 | 0.105 | -0.059*** | 0.000 | 4101 |
| *Employment* | | | | | |
| Any IGA in past month (%) | 0.594 | 0.619 | 0.025 | 0.182 | 4101 |
| Any non-farm IGA in past month (%) | 0.266 | 0.307 | 0.041** | 0.012 | 4101 |
| Earnings in past month (NRs) | 1201.970 | 1295.522 | 93.552 | 0.285 | 4069 |
| Earnings > 3000 in past month (%) | 0.172 | 0.197 | 0.025* | 0.094 | 4101 |
| Trade-specific IGA in past month (%) | 0.154 | 0.189 | 0.035** | 0.014 | 4101 |
| Hours worked past month | 62.774 | 71.502 | 8.728*** | 0.008 | 4101 |
| *Empowerment* | | | | | |
| Any savings (%) | 0.585 | 0.604 | 0.019 | 0.311 | 4080 |
| Total Cash Savings (NRs) | 3114.676 | 3177.379 | 62.703 | 0.832 | 4080 |

*Notes*: This table reports average values for treatment and control groups, with p-value of a Student's t-test for equality of means between the two groups. The tests are conducted on the panel sample (those interviewed at baseline and follow-up). Standard errors are clustered by training course. “ITT” indicates that treatment is defined as having a score that qualifies the respondent for an EF training course.
*** Significant at the 1 percent level.
** Significant at the 5 percent level.
* Significant at the 10 percent level.

**TABLE B2: TAKE-UP OF EF TRAINING (YEAR AFTER BASELINE SURVEY)**

| | **Participated in an EF training course** | | **Did not participate in an EF training course** | |
|---|---|---|---|---|
| | Number | Percent | Number | Percent |
| 2010 Cohort (N=1372) | | | | |
| Assigned to Treatment (N=1040) | 671 | 64.52% | 369 | 35.48% |
| Assigned to Control (N=332) | 86 | 25.90% | 246 | 74.10% |
| 2011 Cohort (N=1415) | | | | |
| Assigned to Treatment (N=1110) | 826 | 74.41% | 284 | 25.59% |
| Assigned to Control (N=305) | 110 | 36.07% | 195 | 63.93% |
| 2012 Cohort (N=1306) | | | | |
| Assigned to Treatment (N=889) | 597 | 67.15% | 292 | 32.85% |
| Assigned to Control (N=417) | 127 | 30.46% | 290 | 69.54% |

*Notes*: There are four individuals from the 2011 cohort and five individuals from 2010 whose status in the EF database is unknown. For these individuals, we rely on the respondent's self-report of whether they took an EF training in the past year for the ATT results. The table only includes those individuals who were surveyed for the first follow-up.

**TABLE B3: FIRST-STAGE PROPENSITY SCORE ESTIMATES**

| **Dependent variable:** | **Treat (ITT)** |
|---|---|
| Age of applicant | 0.000 |
| | (0.009) |
| Sex of applicant (1=Female) | 0.056 |
| | (0.094) |
| Education level of applicant | -0.010 |
| | (0.010) |
| Education of household head | -0.003 |
| | (0.006) |
| Household size | 0.007 |
| | (0.007) |
| Married (1=Yes) | -0.038 |
| | (0.074) |
| Has child (1=Yes) | 0.260** |
| | (0.103) |
| Number of children | -0.079** |
| | (0.039) |
| Any IGA (1= Yes) | 0.057 |
| | (0.072) |
| Zero earnings (1=Yes) | -0.075 |
| | (0.069) |
| Janajati (1=Yes) | 0.031 |
| | (0.443) |
| Dalit (1=Yes) | 0.436 |
| | (0.690) |
| Muslim (1=Yes) | 4.127*** |
| | (0.357) |
| Analytical Ability (0-5) | -0.011 |
| | (0.019) |
| Entrepreneurial score (0-32) | -0.003 |
| | (0.004) |
| Financial literacy (1=Yes) | -0.047 |
| | (0.050) |
| N | 4449 |
| Pseudo $R^2$ | 0.050 |

*Notes*: Standard errors (reported in brackets) clustered by training course. "Treat (ITT)" equals 1 if individual qualified for a training course and 0 otherwise. Other independent variables (not shown): district and T&E provider fixed effects, training-type categories, quintiles of household wealth. All variables measured at baseline. Although baseline data were collected on 4,677 individuals, incomplete data on ethnicity reduces the number of observations to 4,449.
*** Significant at the 1 percent level.
** Significant at the 5 percent level.
* Significant at the 10 percent level.

## TABLE B4: EMPLOYMENT (ITT), 2010-2012 POOLED COHORTS

| | Baseline mean | OLS (1) | IPSW (2) | NN (3) |
|---|---|---|---|---|
| Any IGA (1=Yes) | 0.612 | 0.071*** | 0.093*** | 0.070*** |
| | [0.487] | (0.022) | (0.022) | (0.020) |
| Any non-farm IGA (1=Yes) | 0.296 | 0.149*** | 0.160*** | 0.150*** |
| | [0.457] | (0.023) | (0.024) | (0.021) |
| Trade-specific IGA (1=Yes) | 0.18 | 0.182*** | 0.187*** | 0.184*** |
| | [0.384] | (0.023) | (0.025) | (0.020) |
| Hours worked in past month | 69.261 | 18.740*** | 21.130*** | 19.014*** |
| | [87.273] | (3.890) | (4.148) | (3.940) |
| Earnings | 1271.542 | 856.087*** | 921.323*** | 850.880*** |
| | [2197.669] | (152.941) | (159.517) | (135.139) |
| Logged earnings | 3.291 | 0.957*** | 1.209*** | 0.975*** |
| | [3.817] | (0.191) | (0.203) | (0.173) |
| Earnings > 3000 NRs. (1=Yes) | 0.19 | 0.130*** | 0.140*** | 0.131*** |
| | [0.393] | (0.021) | (0.022) | (0.020) |
| Clustered Standard Errors | | Yes | Yes | No |

*Notes*: All columns report difference-in-difference estimates. "ITT" indicates that everyone whose score qualified them for a given training event is included in the "treatment" group.
Standard errors (reported in brackets) clustered at the event level where possible. Self-employment and location of work were not asked in 2010.
*** Significant at the 1 percent level.
** Significant at the 5 percent level.
* Significant at the 10 percent level.

**TABLE B5: EMPLOYMENT BY TRADE (ITT), 2010-2012 COHORTS**

| | Pooled 2010-2012 Cohorts IPSW Model (ITT Effects) | | | |
|---|---|---|---|---|
| | any nonfarm IGA | trade-specific IGA | monthly earnings (NRs) | earnings > 3000 |
| | (1) | (2) | (3) | (4) |
| Full Sample (pooled across all training types) | 0.160*** | 0.187*** | 921.323*** | 0.140*** |
| | (0.024) | (0.025) | (159.517) | (0.022) |
| Training: Farming (N=92) | 0.155* | -0.059 | 1167.151 | 0.081 |
| | (0.081) | (0.104) | (1000.983) | (0.169) |
| Training: Poultry Technician (N=41) | 0.226 | 0.342*** | 1139.704 | 0.189 |
| | (0.173) | (0.099) | (969.082) | (0.145) |
| Training: Food prep/Hospitality (N=265) | -0.057 | 0.007 | -965.418 | -0.146 |
| | (0.096) | (0.064) | (1048.109) | (0.095) |
| Training: Electrician & Electronics (N=641) | 0.187*** | 0.258*** | 1282.843*** | 0.160*** |
| | (0.044) | (0.058) | (359.255) | (0.054) |
| Training: Handicraft & Incense stick making (N=235) | 0.107 | 0.207*** | 967.311* | 0.129 |
| | (0.082) | (0.075) | (524.717) | (0.094) |
| Training: Construction (N=1128) | 0.067 | 0.100* | 509.836 | 0.089** |
| | (0.054) | (0.058) | (322.866) | (0.038) |
| Training: Beautician/Barber (N=239) | 0.247*** | 0.402*** | 1529.259*** | 0.241*** |
| | (0.094) | (0.089) | (533.151) | (0.078) |
| Training: Weaving/Tailoring/Garment (N=1461) | 0.249*** | 0.222*** | 1185.755*** | 0.196*** |
| | (0.038) | (0.037) | (233.399) | (0.035) |
| Clustered standard errors (by event) | Yes | Yes | Yes | Yes |

*Notes*: No poultry technician trainings were included in the 2011 sample.
*** Significant at the 1 percent level.
** Significant at the 5 percent level.
* Significant at the 10 percent level.

## TABLE B7: EMPLOYMENT OUTCOMES (ITT), BY GENDER, 2010-2012 POOLED COHORTS

**IPSW MODEL**

| | Baseline mean for men | Baseline mean for women | Men | Women | Difference |
|---|---|---|---|---|---|
| | | | (1) | (2) | (3) |
| Any IGA (1=Yes) | 0.774 | 0.518 | 0.025 | 0.130*** | -0.106** |
| | [0.418] | [0.500] | (0.035) | (0.028) | (0.045) |
| Any non-farm IGA (1=Yes) | 0.471 | 0.195 | 0.105** | 0.192*** | -0.087* |
| | [0.499] | [0.396] | (0.044) | (0.028) | (0.051) |
| Trade-specific IGA (1=Yes) | [0.499] | [0.396] | 0.147*** | 0.209*** | -0.062 |
| | 0.295 | 0.113 | (0.046) | (0.028) | (0.053) |
| Hours worked in past month | 107.772 | 46.887 | 11.564 | 26.287*** | -14.723 |
| | [99.126] | [70.525] | (8.796) | (4.242) | (9.795) |
| Total monthly earnings (NRs) | 2137.947 | 774.683 | 681.698** | 1036.088*** | -354.390 |
| | [2539.479] | [1796.025] | (300.488) | (173.214) | (341.802) |
| Logged earnings | 4.796 | 2.428 | 0.281 | 1.688*** | -1.407*** |
| | [3.917] | [3.476] | (0.341) | (0.237) | (0.414) |
| Earnings > 3000 NRs. (1=Yes) | 0.350 | 0.098 | 0.091** | 0.166*** | -0.075 |
| | [0.477] | [0.297] | (0.039) | (0.027) | (0.047) |
| Clustered Standard Errors | | | Yes | Yes | Yes |

*Notes:* Standard errors (reported in brackets) clustered at the event level where possible.
Younger women (aged 16 to 24) compared to older women (age 25 to 35).
*** Significant at the 1 percent level.
** Significant at the 5 percent level.
* Significant at the 10 percent level.

## TABLE B8: EMPLOYMENT (ITT), DISAGGREGATED BY WOMEN, 2010-2012 COHORTS

| | | | IPSW MODEL | | |
|---|---|---|---|---|---|
| | Baseline mean for young women | Baseline mean for women | Young women | Older women | Difference |
| | | | (1) | (2) | (3) |
| Any IGA (1=Yes) | 0.5 | 0.543 | 0.140*** | 0.127*** | 0.013 |
| | [0.500] | [0.498] | (0.040) | (0.041_ | (0.058) |
| Any non-farm IGA (1=Yes) | 0.168 | 0.225 | 0.196*** | 0.187*** | 0.009 |
| | [0.374] | [0.418] | (0.041) | (0.042) | (0.062) |
| Trade-specific IGA (1=Yes) | 0.096 | 0.131 | 0.213*** | 0.204*** | 0.010 |
| | [0.295] | [0.338] | (0.036) | (0.039) | (0.048) |
| Hours worked in past month | 39.569 | 55.560 | 25.348*** | 27.881*** | -2.533 |
| | [62.475] | [78.058] | (5.746) | (6.83) | (9.259) |
| Total monthly earnings (NRs) | 560.537 | 1026.533 | 834.168*** | 1283.426*** | -449.259 |
| | [1438.980] | [2113.341] | (183.918) | (283.993) | (320.888) |
| Logged earnings | 2.063 | 2.857 | 1.633*** | 1.791*** | -0.158 |
| | [3.264] | [3.665] | (0.329) | (0.366) | (0.505) |
| Earnings > 3000 NRs. (1=Yes) | 0.071 | 0.131 | 0.144*** | 0.192*** | -0.048 |
| | [0.256] | [0.337] | (0.030) | (0.043) | (0.051) |
| Clustered Standard Errors | | | Yes | Yes | Yes |

*Notes:* Standard errors (reported in brackets) clustered at the event level where possible.
Younger women (aged 16 to 24) compared to older women (age 25 to 35).
*** Significant at the 1 percent level.
** Significant at the 5 percent level.
* Significant at the 10 percent level.

**TABLE B9: BASELINE BALANCE TESTS 2010-2012 POOLED COHORTS (ATT)**

| | Control | Treatment | Difference | p-value | N |
|---|---|---|---|---|---|
| *Demographics* | | | | | |
| Female (%) | 0.633 | 0.632 | -0.000 | 0.994 | 4101 |
| AGEI (women aged 16-24) (%) | 0.333 | 0.331 | -0.002 | 0.884 | 4101 |
| Dalit (%) | 0.097 | 0.069 | -0.027** | 0.028 | 4037 |
| Janajati (%) | 0.413 | 0.486 | 0.073*** | 0.000 | 4037 |
| Muslim (%) | 0.019 | 0.026 | 0.007* | 0.062 | 4037 |
| Age | 24.389 | 24.268 | -0.121 | 0.593 | 4101 |
| Currently Married (%) | 0.585 | 0.595 | 0.010 | 0.577 | 4101 |
| Any Children (%) | 0.509 | 0.529 | 0.020 | 0.272 | 4101 |
| Completed SLC (10th grade) (%) | 0.147 | 0.101 | -0.046*** | 0.000 | 4101 |
| *Employment* | | | | | |
| Any IGA in past month (%) | 0.593 | 0.625 | 0.032* | 0.082 | 4101 |
| Any non-farm IGA in past month (%) | 0.283 | 0.306 | 0.023 | 0.142 | 4101 |
| Earnings in past month (NRs) | 1258.539 | 1280.494 | 21.955 | 0.788 | 4069 |
| Earnings > 3000 in past month (%) | 0.185 | 0.194 | 0.010 | 0.490 | 4101 |
| Trade-specific IGA in past month (%) | 0.173 | 0.185 | 0.011 | 0.397 | 4101 |
| Hours worked past month | 66.292 | 71.308 | 5.016 | 0.128 | 4101 |
| *Empowerment* | | | | | |
| Any savings (%) | 0.580 | 0.612 | 0.032* | 0.079 | 4080 |
| Total Cash Savings (NRs) | 3246.505 | 3102.511 | -143.994 | 0.608 | 4080 |

*Notes:* This table reports average values for treatment and control groups, with p-value of a Student's t-test for equality of means between the two groups. The tests are conducted on the panel sample (those interviewed at baseline and follow-up). Standard errors are clustered by training course. “ITT” indicates that treatment is defined as having a score that qualifies the respondent for an EF training course.

*** Significant at the 1 percent level.

** Significant at the 5 percent level.

* Significant at the 10 percent level.

**TABLE B10: FIRST-STAGE PROPENSITY SCORES (ATT)**

| **Dependent variable:** | **TREAT** |
|---|---|
| Age of applicant | 0.001 |
| | (0.009) |
| Sex of applicant (1=Female) | 0.026 |
| | (0.086) |
| Education level of applicant | 0.005 |
| | (0.009) |
| Education of hh head | -0.001 |
| | (0.006) |
| Household size | 0.008 |
| | (0.006) |
| Married (1=Yes) | -0.055 |
| | (0.062) |
| Has child (1=Yes) | 0.291*** |
| | (0.086) |
| Number of children | -0.113*** |
| | (0.034) |
| Any IGA (1= Yes) | 0.117* |
| | (0.069) |
| Zero earnings (1=Yes) | -0.034 |
| | (0.062) |
| Janajati (1=Yes) | 0.483 |
| | (0.661) |
| Dalit (1=Yes) | 0.146 |
| | (0.798) |
| Muslim (1=Yes) | 4.865*** |
| | (0.370) |
| Analytical Ability (0-5) | 0.028 |
| | (0.018) |
| Entrepreneurial score (0-32) | -0.004 |
| | (0.004) |
| Financial literacy (1=Yes) | -0.053 |
| | (0.050) |
| N | 4490 |
| Pseudo $R^2$ | 0.071 |

*Notes*: Standard errors (reported in brackets) clustered by training course. "Treat (ATT)" equals 1 if individual participated in a training course and 0 otherwise. Other independent variables (not shown): district and T&E provider fixed effects, training-type categories, quintiles of household wealth. All variables measured at baseline. Although baseline data were collected on 4,677 individuals, incomplete data on ethnicity reduces the number of observations to 4,449.
*** Significant at the 1 percent level.
** Significant at the 5 percent level.
* Significant at the 10 percent level.

**TABLE B11: EMPLOYMENT (ATT), 2010-2012 COHORTS**

| | Baseline mean | OLS<br>(1) | IPSW<br>(2) | NN<br>(3) |
|---|---|---|---|---|
| Any IGA (1=Yes) | 0.612 | 0.085*** | 0.123*** | 0.089*** |
| | [0.487] | (0.021) | (0.021) | (0.018) |
| Any non-farm IGA (1=Yes) | 0.296 | 0.203*** | 0.203*** | 0.206*** |
| | [0.457] | (0.024) | (0.025) | (0.019) |
| Trade-specific IGA (1=Yes) | 0.18 | 0.244*** | 0.233*** | 0.243*** |
| | [0.384] | (0.024) | (0.024) | (0.017) |
| Hours worked in past month | 69.261 | 29.581*** | 31.545*** | 30.484*** |
| | [87.273] | (4.092) | (4.171) | (3.475) |
| Earnings | 1271.542 | 976.240*** | 1099.759*** | 1018.001*** |
| | [2197.669] | (135.081) | (131.653) | (119.605) |
| Logged earnings | 3.291 | 1.392*** | 1.554*** | 1.432*** |
| | [3.817] | (0.195) | (0.195) | (0.152) |
| Earnings > 3000 NRs. (1=Yes) | 0.19 | 0.159*** | 0.168*** | 0.161*** |
| | [0.393] | (0.022) | (0.021) | (0.018) |
| Clustered Standard Errors | | Yes | Yes | No |

*Notes*: All columns report difference-in-difference estimates. "ITT" indicates that everyone whose score qualified them for a given training event is included in the "treatment" group.
Standard errors (reported in brackets) clustered at the event level where possible.
Self-employment and location of work were not asked in 2010.
*** Significant at the 1 percent level.
** Significant at the 5 percent level.
* Significant at the 10 percent level.

**TABLE B12: EMPLOYMENT 2010-2012 POOLED (ATT IPSW MODEL)**

| | any nonfarm IGA | trade-specific IGA | monthly earnings (NRs) | earnings > 3000 |
|---|---|---|---|---|
| | (1) | (2) | (3) | (4) |
| Full Sample (pooled across all training types) | 0.203*** | 0.233*** | 1099.759*** | 0.168*** |
| | (0.025) | (0.024) | (131.653) | (0.021) |
| Training: Farming (N=92) | 0.139** | -0.044 | 529.452 | 0.081 |
| | (0.068) | (0.083) | (547.819) | (0.094) |
| Training: Poultry (N=41) | 0.036*** | 0.184* | -707.959 | -0.085** |
| | (0.011) | (0.110) | (442.371) | (0.040) |
| Training: Food prep/Hospitality (N=265) | 0.164 | 0.139* | 1596.182*** | 0.103 |
| | (0.139) | (0.080) | (581.698) | (0.079) |
| Training: Electrician & Electronics (N=641) | 0.191*** | 0.353*** | 884.423** | 0.140*** |
| | (0.064) | (0.054) | (373.850) | (0.047) |
| Training: Handicraft & Incense stick making (N=235) | 0.047 | 0.162** | 1594.242*** | 0.178** |
| | (0.081) | (0.073) | (450.527) | (0.073) |
| Training: Construction (N=1128) | 0.082** | 0.079* | 702.008** | 0.099** |
| | (0.040) | (0.041) | (298.111) | (0.045) |
| Training: Beautician/Barber (N=239) | 0.459*** | 0.501*** | 1966.409*** | 0.282*** |
| | (0.108) | (0.114) | (402.549) | (0.069) |
| Training: Weaving/Tailoring/Garment Making (N=1461) | 0.303*** | 0.306*** | 1272.605*** | 0.235*** |
| | (0.039) | (0.038) | (180.795) | (0.034) |
| Clustered standard errors (by event) | Yes | Yes | Yes | Yes |

*Note:* No poultry technician trainings were included in the 2011 or 2012 samples.
*** Significant at the 1 percent level.
** Significant at the 5 percent level.
* Significant at the 10 percent level.

**TABLE B13: EMPLOYMENT (ATT-IPSW MODEL), BY GENDER, 2010-2012 COHORTS**

| | Baseline mean for men | Baseline mean for women | Men | Women | Difference between men and women |
|---|---|---|---|---|---|
| | [Std Dev] | [Std Dev] | (1) | (2) | (3) |
| Any IGA (1=Yes) | 0.774 | 0.518 | 0.091*** | 0.143*** | -0.051 |
| | [0.418] | [0.500] | (0.030) | (0.028) | (0.040) |
| Any non-farm IGA (1=Yes) | 0.471 | 0.195 | 0.128*** | 0.247*** | -0.120** |
| | [0.499] | [0.396] | (0.041) | (0.030) | (0.050) |
| Trade-specific IGA (1=Yes) | 0.295 | 0.113 | 0.170*** | 0.269*** | -0.098** |
| | [0.456] | [0.317] | (0.038) | (0.030) | (0.048) |
| Hours worked in past month | 107.772 | 46.887 | 23.887*** | 35.808*** | -11.921 |
| | [99.126] | [70.525] | (7.103) | (4.758) | (8.147) |
| Total monthly earnings (NRs) | 2137.947 | 774.683 | 898.782*** | 1185.170*** | -286.387 |
| | [2539.479] | [1796.025] | (255.902) | (141.973) | (286.878) |
| Logged earnings | 4.796 | 2.428 | 0.582* | 2.072*** | -1.490*** |
| | [3.917] | [3.476] | (0.300) | (0.236) | (0.377) |
| Earnings > 3000 NRs. (1=Yes) | 0.350 | 0.098 | 0.115*** | 0.197*** | -0.081* |
| | [0.477] | [0.297] | (0.035) | (0.025) | (0.042) |
| Clustered Standard Errors | | | Yes | Yes | Yes |

*Notes*: Standard errors (reported in brackets) clustered at the event level where possible.
*** Significant at the 1 percent level.
** Significant at the 5 percent level.
* Significant at the 10 percent level.

**TABLE B14: EMPLOYMENT (ATT-IPSW MODEL), FOR WOMEN, 2010-2012 COHORTS**

| | Baseline mean for young women | Baseline mean for women | Young women | Older women | Difference |
|---|---|---|---|---|---|
| | [Std Dev] | [Std Dev] | (1) | (2) | (3) |
| Any IGA (1=Yes) | 0.500 | 0.543 | 0.137*** | 0.159*** | -0.022 |
| | [0.500] | [0.498] | (0.039) | (0.037) | (0.051) |
| Any non-farm IGA (1=Yes) | 0.168 | 0.225 | 0.246*** | 0.252*** | -0.006 |
| | [0.374] | [0.418] | (0.039) | (0.041) | (0.052) |
| Trade-specific IGA (1=Yes) | 0.096 | 0.131 | 0.265*** | 0.276*** | -0.011 |
| | [0.295] | [0.338] | (0.036) | (0.037) | (0.042) |
| Hours worked in past month | 39.569 | 55.560 | 30.988*** | 42.010*** | -11.022 |
| | [62.475] | [78.058] | (6.019) | (7.538) | (9.525) |
| Total monthly earnings (NRs) | 560.537 | 1026.533 | 1063.272*** | 1343.550*** | -280.278 |
| | [1438.980] | [2113.341] | (172.773) | (218.104) | (263.357) |
| Logged earnings | 2.063 | 2.857 | 2.028*** | 2.187*** | -0.159 |
| | [3.264] | [3.665] | (0.301) | (0.325) | (0.406) |
| Earnings > 3000 NRs. (1=Yes) | 0.071 | 0.131 | 0.165*** | 0.235*** | -0.071* |
| | [0.256] | [0.337] | (0.030) | (0.036) | (0.040) |
| Clustered Standard Errors | | | Yes | Yes | Yes |

*Notes:* Standard errors (reported in brackets) clustered at the event level where possible.
Younger women (aged 16 to 24) compared to older women (age 25 to 35).
*** Significant at the 1 percent level.
** Significant at the 5 percent level.
* Significant at the 10 percent level.